\documentclass[manuscript,nonacm]{acmart}
\usepackage[T1]{fontenc}  
\usepackage{libertine} 

\usepackage{enumitem}
\usepackage{tikz}
\usepackage{array}
\usepackage{wrapfig}
\usepackage{booktabs}

\usetikzlibrary{external, shapes.geometric, arrows}
\usepackage{xspace}
\acmDOI{}          
\acmISBN{}
 \title[short]{full}
\usepackage{xcolor}
\usepackage{soul}         
\definecolor{quotegrey}{RGB}{242,242,242}
\sethlcolor{quotegrey}
\newcommand{\pquote}[1]{\hl{\textit{"#1"}}}

\AtBeginDocument{%
  \providecommand\BibTeX{{%
    \normalfont B\kern-0.5em{\scshape i\kern-0.25em b}\kern-0.8em\TeX}}}

\begin{document}

\tikzstyle{5_box_node} = [
    rectangle,
    rounded corners, 
    minimum width=2cm, 
    minimum height=1cm,
    text centered,
    text width = 2.5cm,
    draw=black,
]
\tikzstyle{3_box_node} = [
    rectangle,
    rounded corners, 
    minimum width=3cm, 
    minimum height=1cm,
    text centered,
    text width = 4cm,
    draw=black,
]
\tikzstyle{4_box_node} = [
    rectangle,
    rounded corners, 
    minimum width=3cm, 
    minimum height=1cm,
    text centered,
    text width = 3.2cm,
    draw=black,
]
\tikzstyle{arrow} = [thick,->,>=stealth]

\newcommand{\yaqing}[1]{\textcolor{violet}{#1}}
\newcommand{\sssec}[1]{\vspace*{0.05in}\noindent\textbf{#1}}

\newcommand{\sys}{\text{Scanvas}\xspace}
\newcommand{\keyword}{\text{synergistic opportunity}}
\newcommand{\keywordplu}{\text{synergistic opportunities}}
\newcommand{\keywordcap}{\text{Synergistic Opportunities}}

\title{\sys: Discovering and Developing Synergistic Opportunities in Generative Design Spaces}

\author{Yaqing Yang}
\authornote{The work is done during the internship at Toyota Research Institute.}
\affiliation{%
  \institution{Carnegie Mellon University}
  \city{Pittsburgh, PA}
  \country{USA; }
\institution{Toyota Research Institute}
  \city{Los Altos, CA}
  \country{USA}}
\email{yaqingyy@andrew.cmu.edu}

\author{Aniket Kittur}
\affiliation{%
  \institution{Carnegie Mellon University}
  \city{Pittsburgh, PA}
  \country{USA}}
\email{nkittur@cs.cmu.edu}

\author{Hongyu Howie Wang}
\affiliation{%
  \institution{Carnegie Mellon University}
  \city{Pittsburgh, PA}
  \country{USA}}
\email{howiewang@cmu.edu}

\author{Nikolas Martelaro}
\affiliation{%
  \institution{Carnegie Mellon University}
  \city{Pittsburgh, PA}
  \country{USA}}
\email{nikmart@cmu.edu}

\author{Matt Klenk}
\affiliation{%
  \institution{Toyota Research Institute}
  \city{Los Altos, CA}
  \country{USA}}
\email{matt.klenk@tri.global}

\author{Yan-Ying Chen}
\affiliation{%
  \institution{Toyota Research Institute}
  \city{Los Altos, CA}
  \country{USA}}
\email{yan-ying.chen@tri.global}

\author{Matthew K. Hong}
\affiliation{%
  \institution{Toyota Research Institute}
  \city{Los Altos, CA}
  \country{USA}}
\email{matt.hong@tri.global}

\renewcommand{\shortauthors}{Yang et al.}
\begin{teaserfigure}
 \includegraphics[width=\linewidth]{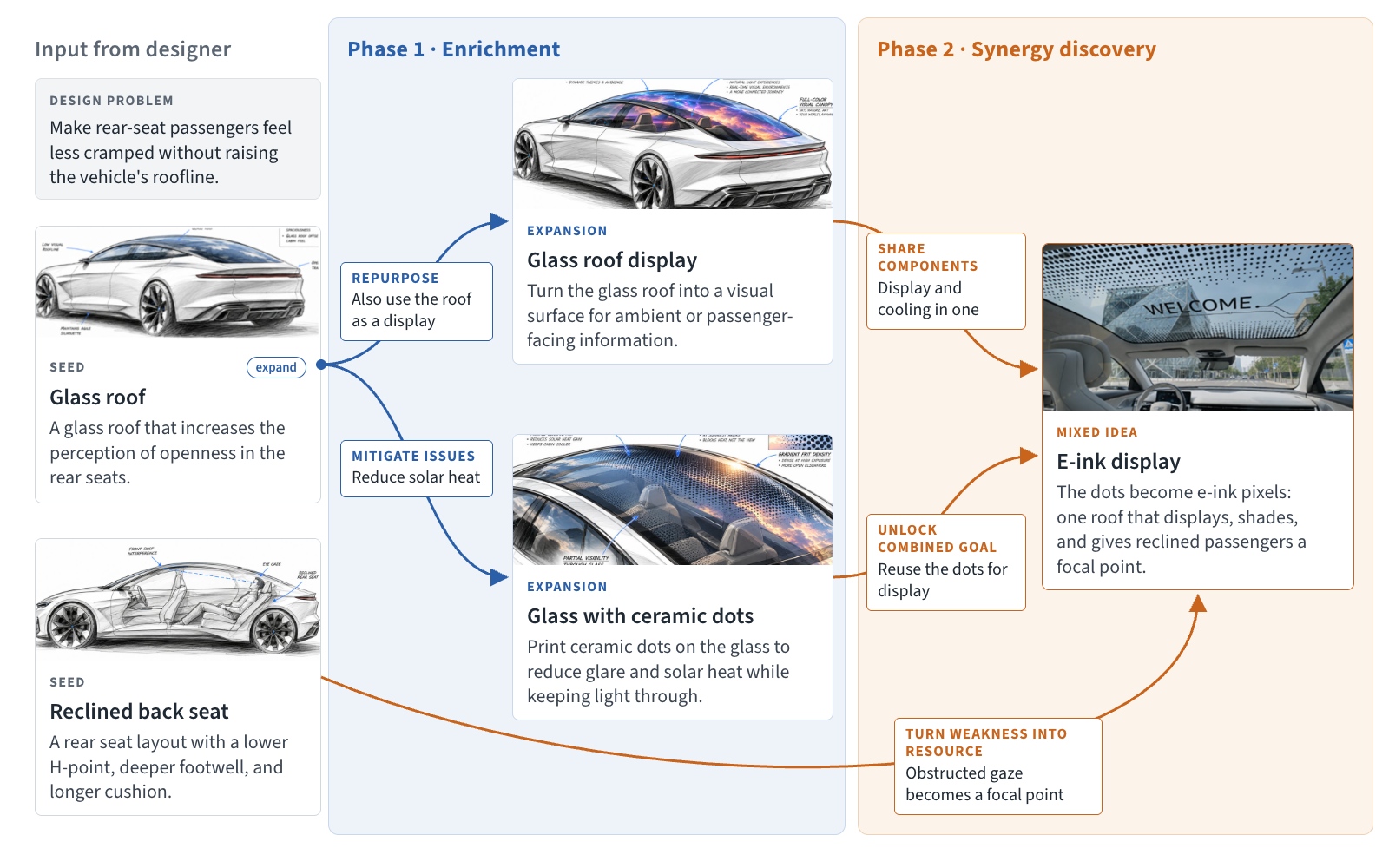}
  \caption{\sys supports discovering and developing synergistic opportunities that create ideas with greater value and the same or fewer resources through two phases:
\textcircled{1} \sys first enriches the idea space by analyzing each input idea in terms of its behaviors, surplus capabilities, and issues to expose potentially useful properties beyond its original formulation. It then expands each idea into different variants by repurposing existing components, leveraging byproducts, or mitigating issues.
\textcircled{2} \sys then searches the enriched idea space to discover and develop synergistic opportunities using three strategies: combining ideas to unlock new goals, turning weaknesses into resources, and sharing components across functions.}
  \label{fig:pipeline}
  \Description{This teaser figure demonstrates how two seed solution ideas (a glass car roof and a reclined back seat) can be enriched and synergized into more valuable design concepts (e.g., a glass roof e-ink display that can be viewed by a reclined passenger) to solve a design problem (less claustrophobic feel in the rear vehicle cabin without changing the roofline). The process is fully described in the captions and main text.}
\end{teaserfigure}



\begin{abstract}
Good design is often synergistic, creating super-additive value by linking goals so that existing resources produce greater outcomes. However, finding these synergistic opportunities in sparse design spaces is difficult, and current LLM-supported ideation tools largely default to additive paradigms such as feature blending, variant generation, or local patching. We present \sys, an AI-supported system for systematically discovering and developing synergistic design opportunities. \sys operationalizes synergy through a two-step computational process: first, it decomposes seed ideas into explicit properties—\emph{components}, \emph{behaviors}, \emph{surpluses}, and \emph{issues}—to enrich the design space; second, it systematically searches across enriched ideas using three theory-grounded strategy operators: \textbf{unlocking or strengthening goals}, \textbf{turning weaknesses into resources}, and \textbf{sharing components across functions}. We instantiate \sys as an auto-generation pipeline and an interactive system. Pipeline ablations and a user study with 12 professional designers demonstrate that \sys enables users to surface and develop significantly higher-quality, synergistic concepts compared to LLM ideation baselines.
\end{abstract}





\maketitle

\section{Introduction}
\label{intro}


Good design is often synergistic~\cite{boden1998creativity}. For example, regenerative braking made slowing a vehicle and extending battery range synergistic by repurposing an electric motor as generator, recovering energy that would otherwise be lost. Such synergistic design moves are superadditive, creating greater value from the same or fewer resources, linking goals that were served separately or counter to each other ~\cite{fuller1982synergetics,ulrich1990function}. 

However, finding valuable synergies can be challenging. One historic reason is that fixation on the obvious capabilities of a design’s components can make it difficult to consider the capabilities or opportunities latent in existing components~\cite{duncker1945problem,mccaffrey2012innovation,jansson1991design,crilly2015fixation}.
In the example above, an electric motor's surplus capabilities---such as running in reverse as a generator---need to be considered before those capabilities can be linked to other components and goals, such as capturing energy from the brakes. Latent capabilities, issues, and goals represent a rich but sparse design space of potential synergies that can be difficult to enumerate and traverse~\cite{chan2015importance,wilkenfeld2001similarity,han2018combinator}. 

As a result existing tools, including recent LLM-based systems for supporting ideation, have rarely grappled with the challenges of finding the sparse synergistic opportunities in a large design space. Instead, additive methods of design have been the subject of significant development, including generating many variants of an idea~\cite{difede2022ideamachine}; defining structured parameter spaces to search within~\cite{suh2024luminate}; surfacing branching trade-offs and mitigations~\cite{yang2025flexmind}; or merging ideas by direct prompting, facet recombination, or blending~\cite{difede2022ideamachine,pu2025ideasynth,wang2026nexusai,wang2023popblends,schorlemmer2014coinvent,eppe2018computational}.

In order to better support synergistic design with LLMs, we draw on theory and practice from design, which offers many, partially overlapping, accounts of how synergistic value arises ranging from Altshuller's focus on resolving contradictions through TRIZ's many strategies (e.g., turning a weakness into a strength) to Ulrich's reduction of components through function sharing ~\cite{altshuller1984creativity,hatchuel2009ck,sarasvathy2001causation,ulrich1990function}. Building on this foundation, rather than treating combination as the integration of ideas that are merely compatible, relevant, or novel together, we ask what kinds of relationships allow ideas to strengthen one another. We consolidate previous work into three strategies based on the central idea of linking goals so that existing resources create additional value: adding a goal that existing resources can serve; linking independent or opposing goals to serve each other; or serving multiple goals with fewer shared resources. 



Applied to the glass roof, which we use as a running example (Table~\ref{tab:synergy_strategies_auto}), unlocking a new goal might suggest using the shading dots on the roof as a dot-based e-ink display, so the roof gains a passenger-facing display while keeping the cabin cool. Turning a weakness into a resource could involve leveraging the excess sunlight that heats the cabin for generating electricity for a ventilation fan, so more power for cooling is naturally supplied the greater the need for it. Sharing components could be done by having HVAC perforations built into the roof itself, so the roof distributes cooling air and the low-mounted vent that occupied foot space is no longer needed.

In this work we take a step towards systematizing the exploration of synergistic opportunities in design. To instantiate and evaluate this vision, we present \textbf{\sys}, an AI-assisted creativity support tool to help designers find and develop \emph{synergistic opportunities} for a given problem and seed idea. \sys first decomposes an idea into its properties, including its \emph{components}, \emph{behaviors}, \emph{surpluses}, and \emph{issues}. It then enriches the idea space by developing these properties though repurposing existing components, leveraging byproducts, and mitigating issues. Rather than simply producing more variants, this step expands the idea space with new roles, resources, and relationships, which makes synergistic opportunities easier to find. \sys then searches across the original and enriched ideas to identify relationships that can \textbf{combine components to unlock new goals}, \textbf{turn weaknesses into resources}, or \textbf{share components across functions}. 

We evaluate \sys through both an auto-generation pipeline and an interactive system supporting synergistic search and designer-guided development of promising opportunities. The auto-generation pipeline directly expands the seed ideas of a design problem, searches the enriched idea space for synergistic opportunities, and creates synergistic idea combinations. The interactive system built on this pipeline allows designers to iteratively add, expand, and combine ideas while \sys analyzes the evolving idea space and surfaces potential synergies. An ablation study on the auto-generation pipeline versus LLM-driven controls revealed that both enrichment and synergistic search were both important drivers of synergistic idea quality. An evaluation of the interactive system in a within-subject study with 12 professionals against an LLM-supported ideation baseline suggested that \sys supported the identification and development of synergies whose added value came from how the ideas worked together, rather than from simply accumulating features. Our work makes three contributions:
\begin{itemize}
    \item An end-to-end auto-generation approach that systematically searches for synergistic opportunities and develops idea combinations that create values beyond the original ideas;
    \item \sys, an interactive system based on the above pipeline that supports people in exploring synergies in design space; and
    \item Empirical evidence from a generation-pipeline evaluation and a user study examining how AI system could support synergy finding.
\end{itemize}
\section{Related Work and Design Goal}
We review prior work on three challenges for synergistic ideation: developing ideas to expose new opportunities, identifying relationships through which synergy can arise, and computationally discovering and developing those relationships across ideas. These challenges respectively motivate three design goals for computationally supporting synergistic ideation.

\subsection{Developing Ideas to Prepare for Synergy}

Synergy may not be visible in the original form of an idea, so one or both ideas may need to be further developed before a useful interaction becomes possible. Prior computational systems provide several ways to develop existing ideas. Some apply transformations or generate variations of an idea~\cite{difede2022ideamachine}. Others use structured representations to expose different facets that can be changed or explored: IdeaSynth decomposes research ideas into facets that can be independently developed and recomposed~\cite{pu2025ideasynth}, while NexusAI decomposes ideas into functional structures and supports abstraction and transformation across these structures~\cite{wang2026nexusai}.  Other systems improve ideas by identifying and addressing their limitations~\cite{jiang2024autotriz,yang2025flexmind}. For example, FlexMind analyzes trade-offs in an idea and generates mitigations that allow the idea to be further developed~\cite{yang2025flexmind}. These approaches show that properties beyond an idea's initial solution description---including its structure, abstractions, and issues---can provide useful material for further development.

However, these methods generally use idea development to produce a better, different, or more deeply explored version of an individual idea. The developed variants are not explicitly generated to interact with other ideas. In particular, they provide less systematic support for asking which latent properties should be developed because they could later participate in synergy.

\textbf{Design Goal 1: Enrich ideas to create more opportunities for synergy.}
Our second goal is to systematically develop latent properties of existing ideas so that more components, resources, byproducts, and resolved issues become available for later synergy discovery.

\subsection{Theoretical Strategies for Finding Synergistic Solutions}
\label{prior_work_theory}
Prior design and creativity research suggests several ways to change the relationship between existing design elements, which can lead to synergy. Across this body of literature, we identify three strategies that are particularly relevant to early-stage ideation: \emph{creating or strengthening goals}, \emph{turning weaknesses into resources}, and \emph{sharing components across functions}. Rather than simply combining features, these strategies describe relationships through which a design can create additional value.

\textbf{Combining to Unlock New Goals.}
One form of synergy occurs when existing resources enable a goal that was not previously considered, or allow an existing goal to be better achieved. C--K (concept--knowledge) theory describes \emph{expansive partition}, in which introducing a previously unconsidered property expands what can be designed~\cite{hatchuel2009ck}. \emph{Effectuation} similarly starts from available means and asks what new effects can be created from them, rather than holding goals fixed~\cite{sarasvathy2001causation}. In design practice, designers can also discover unintended properties in their own representations and use them to formulate new requirements~\cite{suwa2000unexpected}. These perspectives suggest that synergy can arise when existing elements are connected to new or strengthened goals.

\textbf{Turning weaknesses into resources.}
A second form of synergy comes from changing the role of a weakness rather than simply reducing it. TRIZ (the theory of inventive problem solving) treats contradictions as opportunities for redesign and seeks solutions that remove the trade-off rather than optimize between competing demands~\cite{ilevbare2013review}. Related work on paradoxical thinking shows that treating conflicting demands as simultaneously important can promote creative solutions~\cite{miron2011paradoxical}. Similar patterns appear at the system level when an unwanted output from one process becomes a useful input to another, as in industrial symbiosis and cradle-to-cradle design~\cite{chertow2000industrial,braungart2007cradle}. Together, these approaches show how something initially treated as a liability can become a resource for another part of the design.

\textbf{Sharing components across functions.}
A third form of synergy arises when the same component supports multiple functions. Ulrich and Seering describe this as \emph{function sharing}: a structural element takes on multiple functions that would otherwise require separate elements~\cite{ulrich1990function}. Related work on structure sharing and product architecture similarly treats the mapping between functions and components as a design variable, allowing one component to serve several functions~\cite{chakrabarti2004new,ulrich1995role}. Such designs create additional value not by adding another component, but by changing how existing components contribute to multiple goals.

While these theories describe useful strategies for creating synergy, many remain high-level conceptual frameworks rather than concrete procedures that can be directly applied computationally. Some more procedural frameworks have been computationally operationalized. TRIZ, for example, provides relatively explicit steps and inventive principles, enabling systems such as AutoTRIZ and TRIZ-GPT to automate parts of contradiction identification and solution generation using LLMs~\cite{jiang2024autotriz,chen2024triz}. However, such systems typically operationalize a particular design method like TRIZ. We find less computational support for systematically bringing together multiple forms of synergy---such as discovering new goals, repurposing weaknesses as resources, and sharing components across functions---within a unified process for analyzing and developing early-stage ideas.

\textbf{Design Goal 2: Systematically operationalize strategies for synergy discovery.}
Our goal is to translate the strategies for achieving greater value with the same or fewer resources into systematic procedures for creating synergistic idea combinations, which can be applied to early-stage ideas, rather than relying on a single design method or leaving the strategies as conceptual guidance.

\subsection{Finding and Developing Synergistic Combinations}

Once ideas have been developed, computational systems can support both \emph{finding what to combine} and \emph{constructing the resulting combination}. For finding candidate combinations, systems use different signals to reduce the large number of possible pairs. Some retrieve concepts that are relevant to a target problem~\cite{han2018combinator,chen2019artificial,wang2024scimon}; others search for unusual or underexplored combinations based on semantic distance, novelty, or patterns learned from existing recombinations~\cite{gu2024llms,chen2025structuring,sternlicht2026chimera}. These approaches can identify pairs that are potentially interesting, but the criteria used to select them do not directly indicate whether the ideas have a complementary relationship through which they can create additional value together.
Given selected inputs, prior systems also provide different methods for constructing combinations. Idea Machine uses LLMs to elaborate combinations of user-selected ideas~\cite{difede2022ideamachine}. IdeaSynth and NexusAI use structured representations of ideas to support more coherent recombination of their subparts or facets~\cite{pu2025ideasynth,wang2026nexusai}. Other approaches identify associations or common structure between concepts and use them to construct a blend, as in PopBlends and COINVENT~\cite{wang2023popblends,schorlemmer2014coinvent}. Computational conceptual blending similarly provides formal mechanisms for aligning and integrating structures from multiple inputs~\cite{eppe2018computational}.

These methods provide strong support for both selecting and constructing combinations, but neither step necessarily targets \emph{synergy}. Candidate-selection methods typically ask whether ideas are relevant, novel, distant, or otherwise promising to connect, rather than whether one idea can strengthen another. Combination methods then focus on integrating the selected inputs into a coherent or novel concept, but a coherent combination can still simply preserve or add together the parents' existing functions. More importantly, both stages generally operate on the available ideas as they currently exist, while a strong synergy may only become possible after one of the parent ideas has first been developed differently. 
Our goal is therefore not simply to find unusual pairs or construct coherent combinations, but to use the forms of synergy identified in the previous section as an explicit search and generation target. This means looking for relationships in which ideas can jointly enable or strengthen a goal, turn a weakness into a resource, or share components across functions, and then developing the combination around that relationship.

\textbf{Design Goal 3: Search for and develop synergistic relationships across ideas.}
We want to intentionally search the enriched idea space for complementary relationships and construct combinations in which additional value arises from the interaction between the ideas, rather than only from adding their existing functions together.
\section{System}
\sys aims to turn opportunities within individual ideas into materials for discovering synergies across ideas.
 \sys uses the two-stage LLM-based pipeline shown in Figure~\ref{fig:pipeline}. It first enriches individual ideas by making their behaviors, surplus capabilities, and issues explicit and expanding them into new variants (DG1). It then operationalizes the synergy strategies systematically (DG2) to searches across the enriched idea space for automatically developing synergistic idea combinations (DG3).
We instantiate this approach in two forms to support complementary needs: an and-to-end auto-generation pipeline that enables scalable synergy search with minimal user effort; and an interactive system that allows users to guide the exploration, flexibly develop promising ideas, and pursue synergistic opportunities based on their own judgment.

\subsection{Generation Pipeline}
\label{generation}
\begin{figure}[ht]
  \centering
 \includegraphics[width=\linewidth]{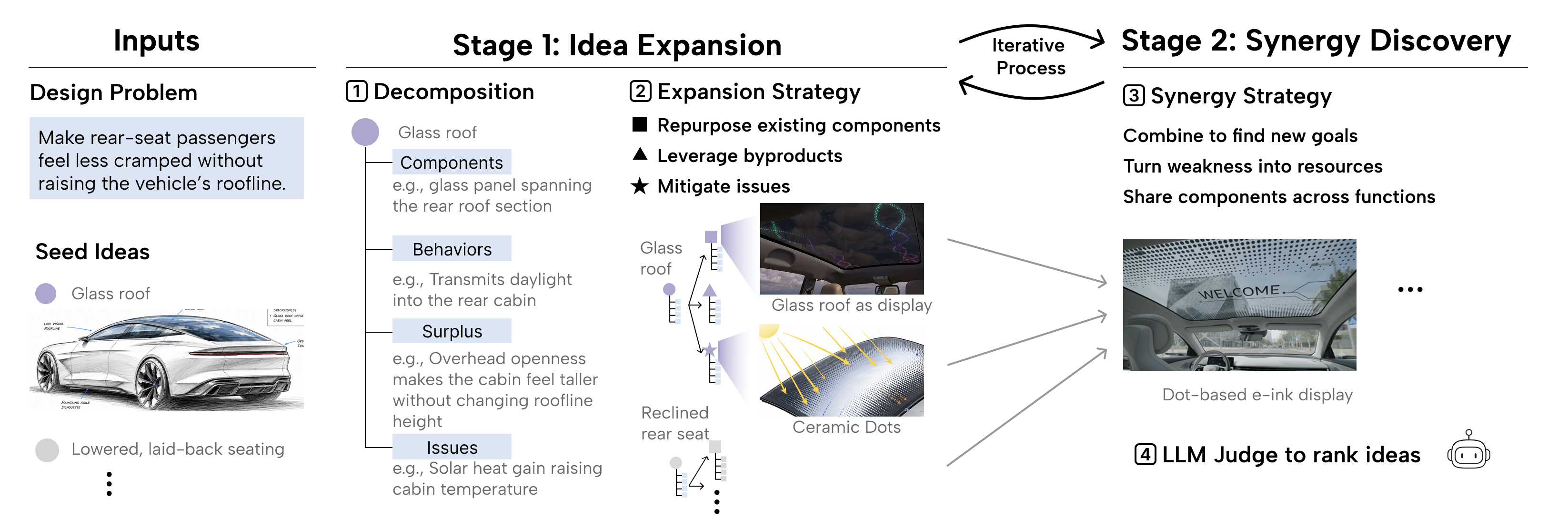}
  \caption{Pipeline for synergy generation. Given a design problem and seed ideas, \textcircled{1} each idea is decomposed into components, behaviors, surplus capabilities, and issues. \textcircled{2} Expansion strategies repurpose components, leverage byproducts, or mitigate issues to generate new ideas. \textcircled{3} Original and expanded ideas are searched for synergies that unlock new goals, turn weaknesses into resources, or share components across functions. \textcircled{4} An LLM judge ranks the resulting ideas to surface promising candidates. Expansion and synergy discovery can be applied iteratively.
}
  \label{fig:auto-pipeline}
  \Description{The figure illustrates the two-stage pipeline for generating synergistic ideas. Starting from a design problem and a set of seed ideas, Stage 1 decomposes each idea into components, behaviors, surplus capabilities, and issues, then applies three expansion strategies—repurposing existing components, leveraging byproducts, and mitigating issues—to generate new directions. In the example, a glass roof is expanded into ideas such as a roof display and the use of ceramic dots. Stage 2 searches across the original and expanded ideas using synergy strategies that combine ideas to unlock new goals, turn weaknesses into resources, or share components across functions. These relationships produce synergistic concepts such as a dot-based e-ink roof display, which are then ranked by an LLM judge. The two stages can be applied iteratively so newly generated ideas can support further expansion and synergy discovery.
}
\end{figure}

The auto-generation pipeline supports scalable synergy discovery with minimal user effort. Given a design problem and seed ideas, it automatically produces synergistic idea combinations by applying the enrichment and synergy discovery stages as in Figure~\ref{fig:auto-pipeline}.

\subsubsection{Idea Enrichment}
\label{method:enrich}
\textbf{To support later synergy discovery, \sys first enriches each idea by structurally decomposing it and expanding it into different variants, so that its components, resources, byproducts, and unresolved issues become available as materials for later synergy construction (DG1).} 
As shown in Figure~\ref{fig:auto-pipeline}(stage 1, \textcircled{1}), rather than directly prompting an LLM to generate variations, \sys decomposes each idea into four dimensions: components, the resources and parts used by the idea~\cite{gero2004situated,bikakis2021repurposing}; behaviors, what those components currently do~\cite{gero2004situated,bikakis2021repurposing}; surplus, underused capabilities or outputs that could support additional uses~\cite{bikakis2021repurposing,chertow2007uncovering}; and issues, remaining problems that could motivate further development~\cite{dorst2001creativity,ilevbare2013review}. This decomposition offers a key advantage: it gives each idea a structured representation of properties that can be surfaced to create new opportunities for systematic idea combination and recombination. Based on this decomposition, \sys applies three expansion strategies drawn from prior literature as shown in Figure~\ref{fig:auto-pipeline}(stage 1, \textcircled{2}). \textit{Repurpose Existing Components}~\cite{bikakis2021repurposing} assigns an existing component an additional function to achieve another task-relevant goal with little additional resource cost. \textit{Leverage Byproducts}~\cite{chertow2007uncovering,bikakis2021repurposing} redirects an otherwise unused output or capability toward a new benefit. \textit{Mitigate Issues}~\cite{dorst2001creativity} develops a variation that directly addresses an identified issue through changes to the current solution. Together, these strategies expand each idea in ways intended to expose additional resources, functions, outputs, and unresolved problems that can provide material for subsequent synergy discovery.

\subsubsection{Synergy Discovery}
\textbf{To search for and develop synergies across ideas at scale (DG3), \sys operationalizes the conceptual synergy strategies (DG2) in an LLM-based pipeline that identifies synergistic idea combinations within the enriched idea space from last step. }
As shown in Figure~\ref{fig:auto-pipeline}(stage 2, \textcircled{3}), \sys uses three complementary synergy rules inspired by prior literature as listed in section~\ref{prior_work_theory}. Table~\ref{tab:synergy_strategies_auto} shows the strategies and their examples. \textit{Combine to Unlock New Goals} searches for pairs that together produce a valuable outcome that neither idea can achieve alone. \textit{Turn Weaknesses into Resources} looks for cases where an issue, side effect, or byproduct of one idea can serve as a useful input to another. \textit{Share Components Across Functions} identifies combinations in which a common component can support multiple functions, reducing redundant parts or steps while preserving the functions of both ideas. For the auto-generation pipeline, since applying these rules across the idea pool grows the candidate set at combinatorial scale, after the two steps, a judging agent ranks the generated synergies to provide a filter to select the strongest candidates, which can be surfaced for further exploration.


\begin{table*}[t]
\caption{Synergy strategies and examples from automotive glass roof design case.}
\label{tab:synergy_strategies_auto}
\centering
\begin{tabular}{p{0.42\textwidth} p{0.52\textwidth}}
\toprule
\textbf{Strategy Definition} & \textbf{Example} \\
\midrule

\textbf{Combine to Unlock New Goals.}
&
\textbf{Glass roof as display + ceramic dots:}
Turn the ceramic-dot pattern into a dot-based e-ink display integrated into the glass roof. The combination enables a dynamic roof display while retaining the familiar dot pattern when the display is inactive.
\\

\midrule

\textbf{Turn Weaknesses into Resources.}
&
\textbf{Excess sunlight + fan energy demand:}
Use excess sunlight entering the glass roof to generate electricity for a ventilation fan, turning an unwanted source of heat and glare into an energy source for cooling.
\\

\midrule

\textbf{[Reduce] Share Components Across Functions.}
&
\textbf{Glass roof + low-mounted rear-seat vent:}
Integrate HVAC passages and perforations into the glass roof so the roof also distributes cooling air. The same roof structure provides openness and ventilation, reducing reliance on a low-mounted vent that occupies rear foot space.
\\

\bottomrule
\end{tabular}
\end{table*}


\subsection{\sys: Interactive System Walkthrough}

\begin{figure}[ht]
   \centering
  \includegraphics[width=\textwidth]{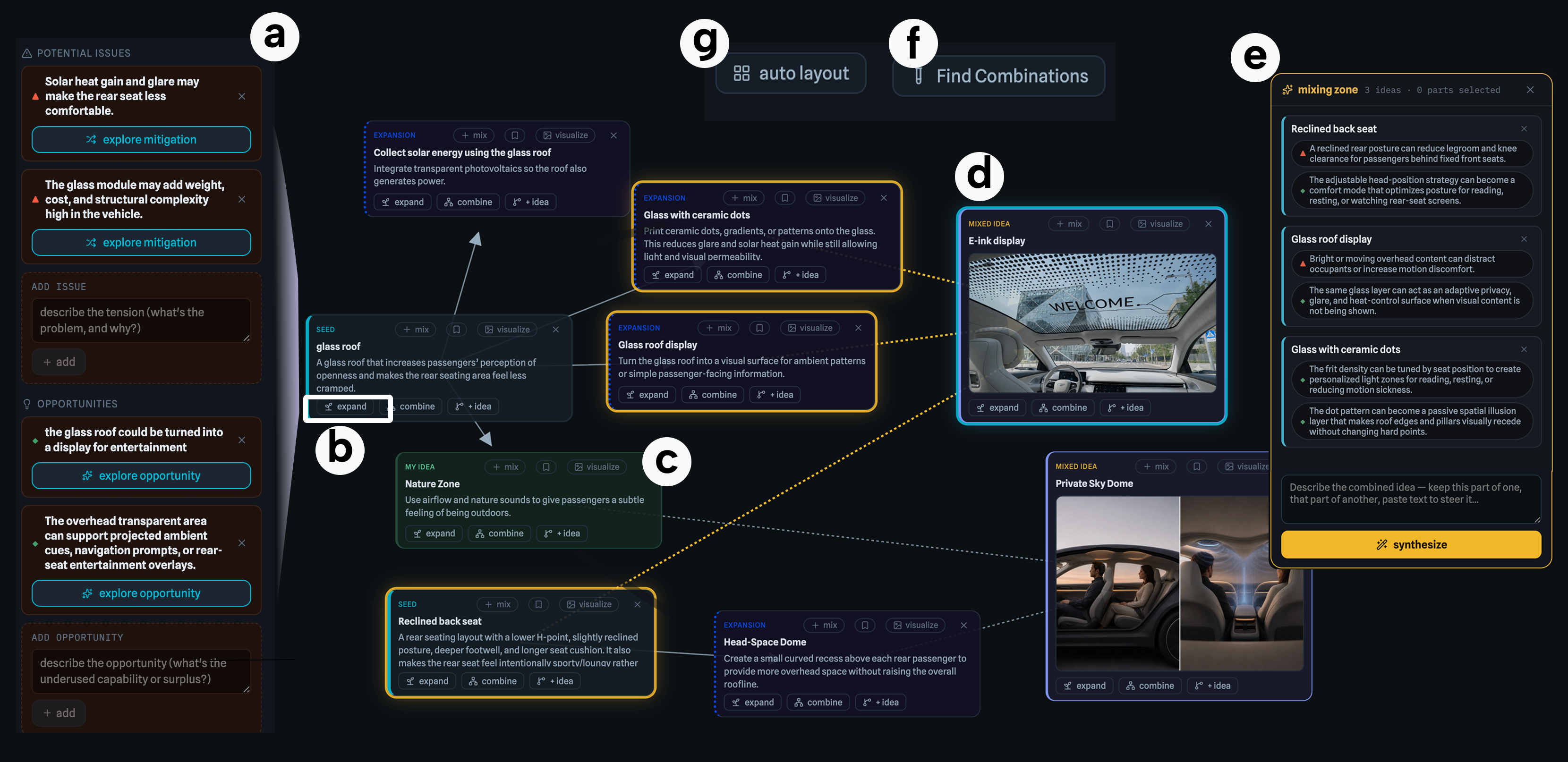}
  \caption{Screenshot of the \sys interface, showing features for (a) decomposing ideas to reveal their properties, (b) expanding ideas into new variants, (e) combining selected ideas using synergy strategies, and (f) automatically discovering synergies across the idea space (g) automatically organizing the canvas }
  \Description{A screenshot of the Scanvas interface. Using the functionalities described in the caption and the main text, a user could start from the seed ideas, expand the ideas, and explore synergy options.}
  \label{fig:teaser}
\end{figure}

The interactive system aims at providing a more flexible way for people to explore synergistic opportunities using the idea enrichment and synergy development strategies. Consider a designer, Sarah, exploring how to make rear-seat passengers feel less cramped without raising the vehicle's roofline. She begins by adding two seed ideas to the canvas: a \textit{glass roof}, which increases visual openness, and a \textit{reclined rear seat}, which provides more usable headroom (Figure~\ref{fig:pipeline}, left). Each idea appears as a card that Sarah can expand or combine with other ideas as shown in Figure~\ref{fig:teaser}. She can also use \textit{Add Idea} (Figure~\ref{fig:teaser} c) on any card to add her own ideas as child nodes, allowing her own thinking and system-generated ideas to develop together on the same canvas.

\subsubsection{Enrich individual ideas.}
Before searching for synergies across ideas, \sys first helps Sarah expose and develop properties within each idea that may become useful in later combinations. By making latent capabilities and issues explicit and expanding individual ideas using the strategies described in Section~\ref{method}, \sys creates additional materials for later synergy discovery.

Selecting the \textit{glass roof} card reveals its issues and surplus capabilities in the sidebar (Figure~\ref{fig:teaser}a). For example, \sys identifies sunlight reaching the roof as a potentially useful resource. Sarah clicks \textit{Explore Opportunity} in the sidebar, generating a \textit{solar-collecting glass roof} that uses this energy to support other cabin functions. The system also identifies glare and solar heat as issues introduced by the glass roof. Clicking \textit{Mitigate Issue} in the sidebar generates a \textit{ceramic-dot roof} (Figure~\ref{fig:pipeline}, middle), where patterned dots provide shading while preserving visual openness. Sarah also clicks \textit{Expand} (Figure~\ref{fig:teaser}b) to generate three variants by repurposing existing components, leveraging byproducts, or mitigating issues. For example, one generated idea is a \textit{glass-roof display} (Figure~\ref{fig:pipeline}, middle), which repurposes the roof from providing visual openness to also serving as a passenger-facing display surface. These features turn latent properties of individual ideas into additional ideas and resources that can support later synergy discovery.

\subsubsection{Combining ideas selected by the user in a synergistic way.}
Building on the enriched idea space, \sys support users to develop synergistic combinations using the ideas they selected. Among the ideas, Sarah sees a promising relationship between the \textit{glass-roof display}, the \textit{ceramic-dot roof}, and the original \textit{reclined rear seat}. Reclining makes the roof more prominent in passengers' field of view; the display gives this surface a new function; and the ceramic dots provide a visual treatment already suited to automotive glass. Sarah clicks \textit{+ Mix} on the three cards to add them to the mix panel (Figure~\ref{fig:teaser} e). There, she can select particular opportunities or issues from the ideas that she wants the combination to focus on, or describe the desired direction in a text box.

Based on her selection, \sys develops an idea \textit{dot-based e-ink roof} with an auto-rendered visualization of the idea. When inactive, the controllable dots resemble conventional ceramic dots and form a subtle pattern matched to the visual style of the vehicle. When activated, the dots rearrange into ambient graphics or simple interface elements for rear passengers. Here, Sarah decides which ideas should be combined, while \sys helps develop their relationship into a coherent concept that creates value beyond the individual ideas.

\subsubsection{Finding synergistic combinations automatically.}
Beyond combinations identified by users, \sys searches across the enriched idea space to discover promising relationships that users might otherwise overlook.
As Sarah continues exploring, the glass roof leads her to think beyond visual openness: if seeing the sky can make passengers feel less enclosed, perhaps other sensations could create a similar feeling. She adds an idea \textit{breeze and nature-sound experience}, which uses airflow and sound to evoke an outdoor environment. In another branch, she adds a \textit{personal head-space dome}, a small curved overhead recess that creates more space above an individual passenger without raising the entire roofline.

Sarah likes the breeze and nature-sound idea but is unsure how to develop it further. She clicks \textit{Combine} on that card, asking \sys to search the canvas for a complementary idea. The system identifies the \textit{personal head-space dome} and combines the two into \textit{Private Sky Domes}. Each passenger receives a slightly raised overhead zone paired with localized airflow and directional nature sounds. The dome creates additional perceived head space, while the localized air and sound create a personalized ``outdoor'' experience without changing the environment for other passengers. 

To explore synergistic opportunities, Sarah clicks \textit{Find Combinations} (Figure~\ref{fig:teaser} f), which searches across all ideas on the canvas for promising pairs. \sys connects the user's breeze idea with the heat and noise problems of the glass roof and proposes a \textit{ventilated double-layer glass roof}. Conditioned air is routed through the cavity between an outer glass layer and a perforated inner layer, cooling the roof before entering the cabin through distributed openings above passengers. The geometry of the inner layer and cavity can also help attenuate noise from the large glass surface. The resulting concept uses a shared structure to address airflow distribution, solar heat, and acoustic comfort together.

\subsubsection{Organizing and browsing inspirations.} 
To support easier exploration of the idea space, \sys automatically generates visualizations for combined ideas and also allows users to visualize any selected idea through a \textit{Visualize} button on each idea card. To reduce cognitive load when navigating many ideas, \sys further provides an \textit{Auto Layout} button (Figure~\ref{fig:teaser}g), which uses text embeddings and UMAP to organize idea cards according to their semantic similarity. Users could also use the bookmark feature on each card to save the ideas they like.

\section{Evaluation Method}
\label{method}
We evaluated \sys in two complementary studies. In Study 1, we evaluated the automated generation pipeline by comparing the full pipeline (Figure~\ref{fig:pipeline}) to two ablated variants to isolate the contributions of the idea enrichment and synergy discovery phases. In Study 2, we conducted a within-subject study with 12 professionals comparing \sys against an LLM-supported ideation baseline. We evaluated both the quality of participants’ resulting ideas through independent expert ratings and their experience with the systems through surveys and semi-structured interview.

\subsection{Study 1: Auto-Generation Evaluation}

We first evaluated whether \sys's idea expansion and synergy-generation strategies could generate good ideas automatically. We compared three generation conditions:

\begin{itemize}
\item \textbf{Full pipeline:} evaluated the complete generation pipeline using both idea expansion and synergy generation, as shown in Figure~\ref{fig:pipeline}. Each of the three seed ideas was first expanded into three variants, producing nine expanded ideas. The synergy-generation procedure considered these expanded ideas together with the three original seed ideas (12 ideas in total) and generated at least 10 synergistic idea combinations. An LLM-based selection procedure then selected the top 10 ideas from the generated set.

\item \textbf{Ablation 1---No expansion:} tested if synergy generation remains effective without first expanding individual ideas. We removed the idea-expansion step and provided the same original seed ideas directly to the synergy-generation procedure, similar as full-pipeline condition.

\item \textbf{Ablation 2---No synergy:} tested whether the benefit comes specifically from \sys's structured synergy-generation strategies rather than from generic idea combination. As in the full pipeline, each of the seed ideas was expanded into three variants. We then replaced \sys's synergy-generation procedure with a direct LLM-based combination step. The LLM received the expanded ideas, the original seed ideas, and the original design problem, and was prompted to ``identify and generate 10  interesting idea combinations.''
\end{itemize}

All three conditions started from the same three seed ideas and produced 10 final ideas per task. 
We evaluated the three conditions on three ideation tasks spanning automotive design, independent cooking support, and accessible museum experiences (Table~\ref{tab:evaluation_tasks}). For each task and condition, the pipeline generated 10 final ideas, resulting in 90 ideas in total ($3$ tasks $\times$ $3$ conditions $\times$ $10$ ideas).

To assess idea quality in relation to our definition of synergy as creating greater value with the same or fewer resources, we selected dimensions that reflect both the idea value and the resources required. Specifically, we assessed four dimensions: \emph{Goal Fulfillment}, how well the idea addresses the original design goal, ensuring that additional benefits remain grounded in the design problem; \emph{New Opportunities}, the additional benefits or opportunities it creates, capturing value beyond the original goal; \emph{Value vs.\ Cost}, whether the value created justifies the required resources or costs, accounting for whether added benefits come with disproportionate demands; and \emph{Interestingness}, whether the idea is compelling and worth further exploration, capturing its promise as a direction for continued design development. We recruited two engineering experts to independently rate each generated idea on these four dimensions using a 7-point scale. One evaluator has ten years of experience in mechanical engineering and is currently a senior engineer in industry. The other has seven years of engineering experience, including full-time experience as a robotics researcher in industry, and is currently a PhD candidate in robotics.


\begin{table*}[t]
\caption{Summary of the design tasks and seed ideas used in the evaluation. 
Contents are summarized from the full task descriptions and seed ideas presented to evaluators. All three tasks were used for the study 1 (auto-generation evaluation), while only Tasks 2 and 3 were used in the study 2 (user study).}
\label{tab:evaluation_tasks}
\centering
\begin{tabular}{p{0.43\textwidth} p{0.51\textwidth}}
\toprule
\textbf{Design Task} & \textbf{Seed Ideas} \\
\midrule

Task 1: Improve physical and perceived rear-seat headroom in a low-roofline vehicle without raising the exterior roof or substantially changing the silhouette.
&
(1) Glass roof for perceived openness; 
(2) reclined backrest positioning the head under a higher roof section; 
(3) head-level airflow creating an open-air sensation.
\\

\midrule

Task 2: Help an older adult with mild cognitive decline follow cooking steps, monitor the stove, and recover from interruptions without a caregiver or complex interface.
&
(1) Stove-side step display; 
(2) inactivity-aware smart burner knob; 
(3) sequentially illuminated ingredient tray.
\\

\midrule

Task 3: Communicate artworks' visual style to visitors with visual impairments through quiet, nonvisual or enhanced-visual methods without altering the artwork.
&
(1) Tactile card encoding composition and brushwork; 
(2) headphones sonifying visual style; 
(3) tablet enhancing visual features for partial vision.
\\

\bottomrule
\end{tabular}
\end{table*}

\subsection{Study 2: User Study of the Interactive System}
We further conducted a within-subjects study to investigate how \sys supports people during interactive ideation compared with an ablated baseline. Each participant completed two ideation tasks, using the full \sys system for one task and the baseline for the other.

\subsubsection{Study Design and Procedure}
Participants followed a fixed procedure in an in-person study, using the same laptop throughout: an introduction, a tutorial on the first system (detailed in the Appendix~\ref{apd:tutorial}), and a 30-minute main task with the first system; they then switched conditions and repeated the tutorial and main task with the second system, followed by a survey and debrief. For each condition, participants were asked to record the ideas they found interesting during the session in a Google Sheet for later assessment. Participants were also asked to think aloud during the timed tasks. 
In each condition, participants worked on either Task 2 or 3 using the same seed ideas listed in Table~\ref{tab:evaluation_tasks}. To reduce potential confounds from task content, fatigue and order effects, we counterbalanced task assignment, condition assignment, and presentation order across the 12 participants.

\subsubsection{Participants}
We recruited 12 participants (7 men and 5 women) through professional networks via Slack and email. We asked participants to complete a pre-survey to collect demographic information and an Alternative Uses Test for later analysis (see survey details in Appendix~\ref{apd:pre-study}). Participants had backgrounds in robotics, electrical engineering, and related engineering fields. They were either full-time engineers in industry ($n=7$) or PhD students in engineering fields ($n=5$), and all had professional industry experience. Participants reported an average of 3.3 years of experience in their current roles ($SD=2.93$). Each study session took approximately 90 minutes, and each participant received a \$50 Amazon gift card as compensation.

\subsubsection{Baseline}
To further isolate the contribution of \sys's structured generation strategies, the baseline used a similar canvas interface, as shown in Figure~\ref{fig:teaser}, but without \sys's automatic synergy-discovery support (Figure~\ref{fig:teaser}f). Users could still use the Expand button to generate idea variants and the +Mix button to select ideas for combination. However, despite the similar front-end interface, the back-end implementation of these two features used zero-shot prompts that simply asked the LLM to ``generate different variants of the idea'' and ``combine the given ideas.'' This baseline therefore provided the basic forms of support that users could typically request from an LLM. Both conditions used GPT-5.5 as the underlying language model.


\subsubsection{Evaluation}
We evaluated the user study results based on both participants' user experience and the outcomes they produced. To assess user experience of the creative process, for both \sys and baseline system, we collected participants' self-reported ratings of perceived feature usefulness and their perceptions of how the systems supported different aspects of the ideation process, including expanding ideas and discovering synergistic idea combinations. For \sys condition, we also examined the standardized Creativity Support Index (CSI)~\cite{cherry2014quantifying}, with the collaboration-related item removed because it was not relevant to our study. Because the two conditions shared a similar interface and we  collected other comparative process measures for both conditions, we omitted the baseline CSI to reduce participants' burden for filling in the surveys. Accordingly, we report the CSI descriptively for \sys rather than using it for direct comparison with the baseline.

To assess idea quality, we collected participants' ratings of their selected ideas from both \sys and baseline system upon completing the user study. We also assessed idea quality through an external panel blind to the conditions and participant information. The same two engineering domain experts, who evaluated the ideas from study 1 for auto-generation results,  independently rated the ideas produced by participants using the same four dimensions: \emph{Goal Fulfillment}, \emph{New Possibilities}, \emph{Value vs.\ Cost}, and \emph{Interestingness}. To assess inter-rater reliability, we calculated the intraclass correlation coefficient (ICC(2,2); two-way random-effects model, absolute agreement, average of two raters) using the ratings from both evaluators across all 201 rated ideas. Reliability was good for all four dimensions: Goal Fulfillment, ICC = .84; New Possibilities, ICC = .73; Value vs.\ Cost, ICC = .73; and Interestingness, ICC = .71, indicating good agreement between the two raters~\cite{koo2016guideline}. 

To examine how broadly participants explored the idea space in each condition, we measured the semantic diversity of the ideas submitted by participants. We followed a common computational approach in creativity research that assesses the diversity of an idea set based on the semantic distances among its ideas~\cite{cox2021directed}. We embedded each idea's text using OpenAI's \texttt{text-embedding-3-small} model~\cite{openai_text_embedding_3_small} and calculated the mean pairwise cosine distance among ideas within each task and condition. We then averaged this measure across the two tasks. Higher values indicate that the ideas within a condition are, on average, more semantically different from one another, and thus that the overall idea set is more semantically diverse. Because this measure is averaged over all pairs of ideas, it is not directly determined by the number of ideas produced in each condition.

\section{Result}
Overall, the results of Study 1, which assessed the auto-generation pipeline, show that ideas produced by the full pipeline received significantly higher ratings than those produced by the ablated baselines (Figure~\ref{fig:rating_full_pipeline}). The results of Study 2, the user study, show that ideas produced in the \sys condition received significantly higher expert ratings than those produced in the baseline condition (Figure~\ref{fig:rating_study}). In terms of user experience, participants also gave higher ratings for how \sys supported idea enrichment and the discovery of synergistic idea combinations compared with the baseline (Figure~\ref{fig:user_rating_process}, left). They also gave \sys high Creativity Support Index ratings, with all dimensions receiving an average score above 5.7 on a 7-point scale (see details in Appendix~\ref{apd:post-study}).

\subsection{Idea Quality}

\begin{figure}[ht]
  \centering
 \includegraphics[width=0.8\linewidth]{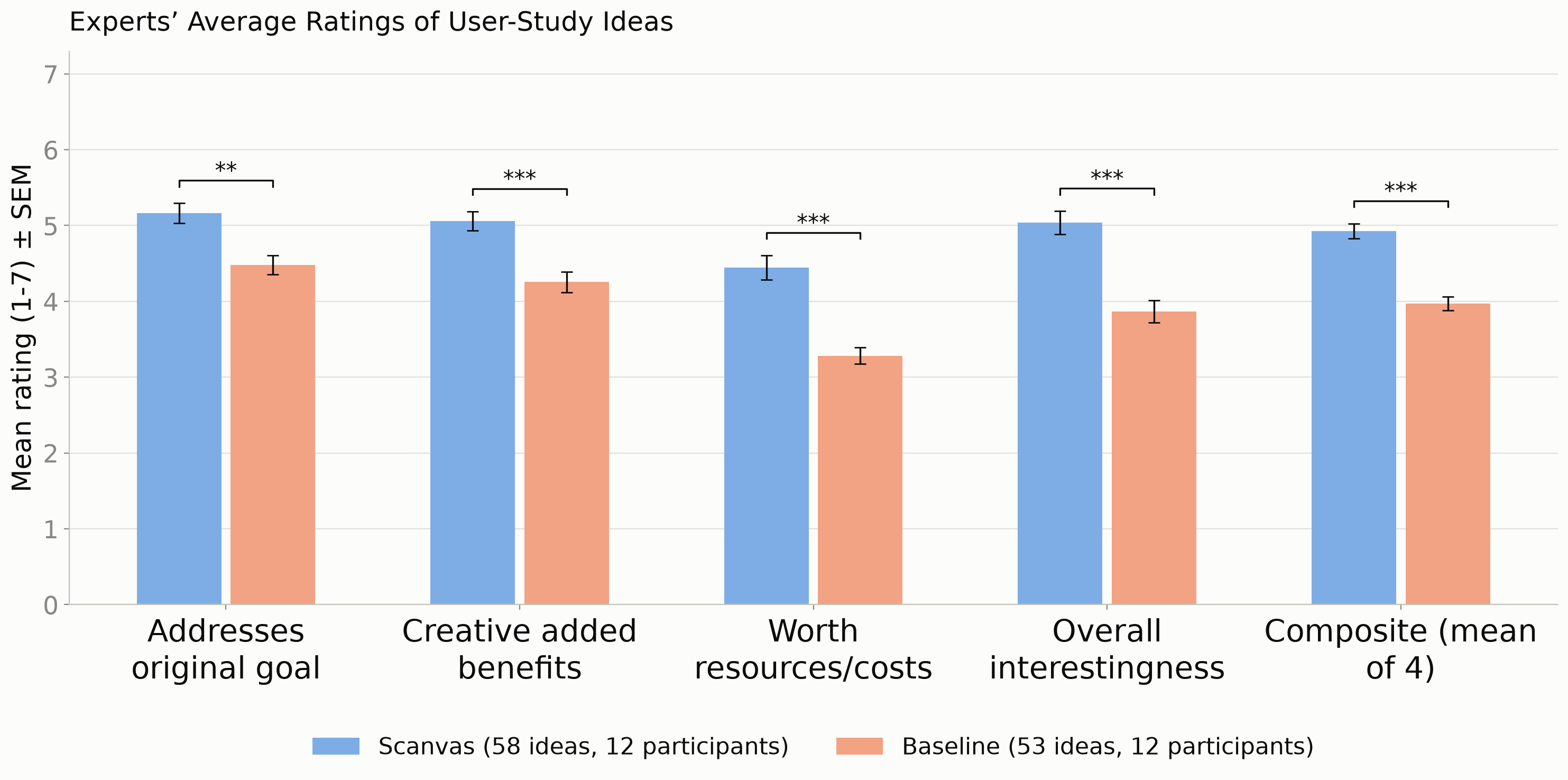}
  \caption{Experts' idea-quality ratings for ideas generated from two conditions in user study. \sys outperforms Baseline across all dimensions.}
  \label{fig:rating_study}
  \Description{Bar charts comparing the average rated quality of ideas generated from the two user study conditions. Scanvas also significantly outperforms the Baseline condition across all dimensions.}
\end{figure}

In the auto-generation evaluation, the ideas from full pipeline received significantly higher ratings than both ablations across all four rating dimensions (Figure~\ref{fig:rating_full_pipeline}). In the user study, ideas produced with \sys also received significantly higher ratings than those produced in the baseline condition across all rating dimensions (Figure~\ref{fig:rating_study}).

\subsubsection{End-to-End Auto-Generation Results}

\begin{figure}[ht]
  \centering
 \includegraphics[width=0.9\linewidth]{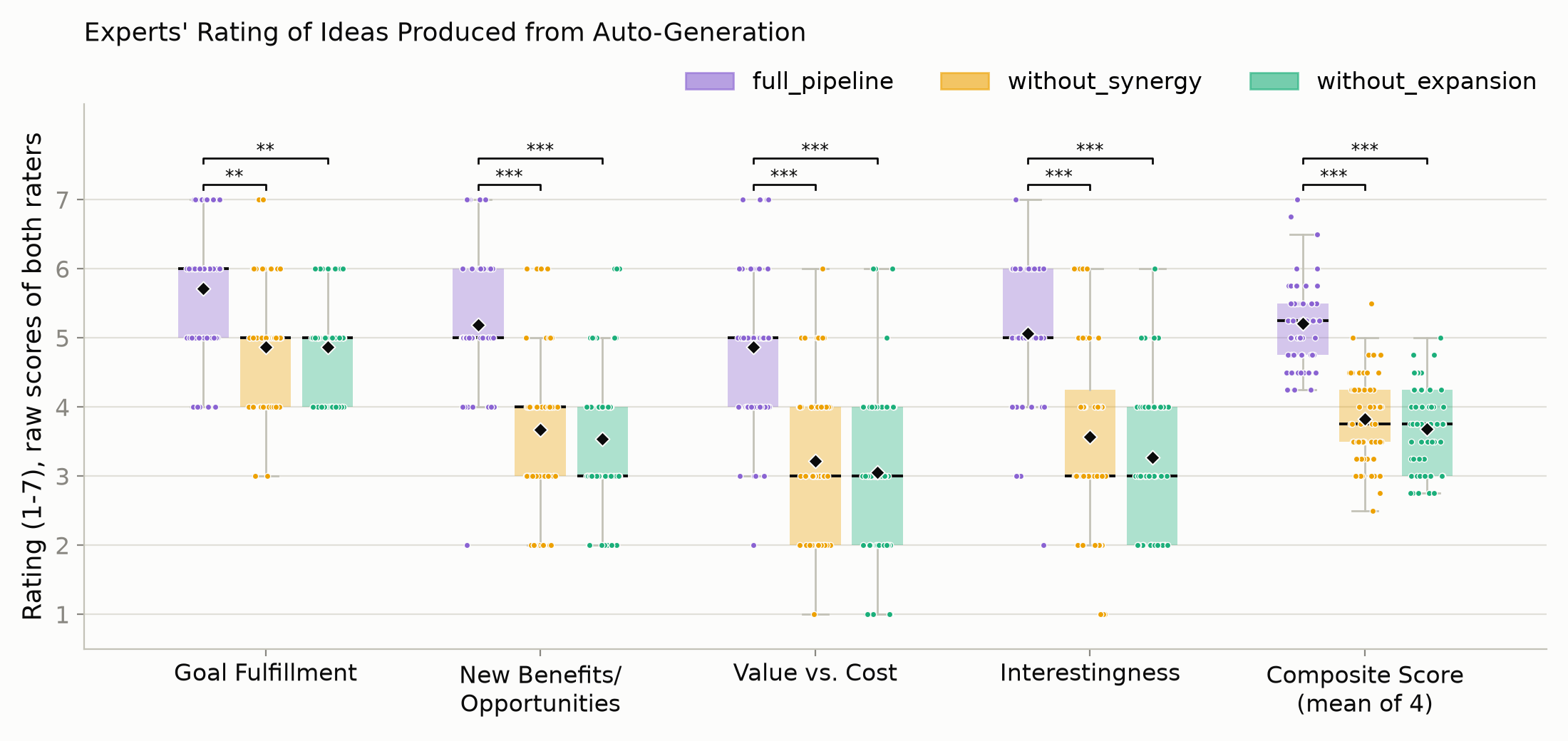}
  \caption{Idea-quality ratings from 2 raters for the end-to-end generation conditions across four dimensions and their composite, pooled over 3 design tasks. The full pipeline outperforms both ablations across all dimensions, while the two ablations do not differ significantly. Points show individual expert ratings; boxes show medians and IQRs, whiskers extend to 1.5 IQR, and diamonds indicate means. Brackets mark significant pairwise differences based on Holm-corrected Mann–Whitney U tests (* \(p<.05\), ** \(p<.01\), *** \(p<.001\)).}
  \label{fig:rating_full_pipeline}
  \Description{Box and whisker plots comparing the rated quality of ideas generated by the three automatic generation methods. Inidividual ratings from the 2 raters for all the idea outputs are also marked on the plots. The full pipeline significantly outperforms both ablations across all dimensions.}
\end{figure}

The full automated pipeline consistently outperformed both ablations (Figure~\ref{fig:rating_full_pipeline}), showing that idea expansion and synergy generation jointly improve idea quality. 
Across all four rating dimensions, 
the largest improvements were in \emph{New Benefits/Opportunities} and \emph{Interestingness}. This pattern aligns with participants' reports in the user study that \sys helped reveal new values in ideas (Figure~\ref{fig:user_rating_process}), providing further evidence that the pipeline improves the LLM's ability to surface new opportunities from the idea space, and produce interesting ideas.

The two ablations did not differ significantly from each other, suggesting that expansion and synergy generation reinforce one another. Expansion develops individual ideas by surfacing additional opportunities and issues, providing richer inputs for identifying useful synergies. In turn, synergy generation connects ideas in ways that explore the opportunities found, and could also reveal further opportunities for expansion. Removing either step may therefore weaken the other. Together, these results suggest that the full pipeline's advantage comes from integrating expansion and synergy generation rather than relying on either process alone.

\subsubsection{User Study Idea Ratings}
Overall, \sys helped participants produce higher-quality ideas than the baseline. Ideas produced in the \sys condition received significantly higher expert ratings on the composite score (\sys: $M=4.93$, $SD=0.76$; baseline: $M=3.99$, $SD=0.66$, $p<.001$). This advantage held across all four rating dimensions, with higher mean ratings for \sys in addressing the original goal (5.16 vs.\ 4.50, $p=.001$), surfacing new benefits (5.06 vs.\ 4.27, $p<.001$), worth relative to resources and costs (4.45 vs.\ 3.28, $p<.001$), and overall interestingness (5.03 vs.\ 3.89, $p<.001$). All comparisons used two-sided Mann--Whitney $U$ tests with Holm correction. Table~\ref{tab:high_low_idea_examples} shows examples of ideas received relatively high and low ratings. The strongest difference was in surfacing \textit{New Benefits/Opportunities}, particularly at the high end of the rating scale. In the \sys condition, 34.5\% of ideas received ratings of 6 or 7 on this dimension, compared with only 3.8\% in the baseline. At the participant level, 8 of 12 participants received higher average idea ratings for this dimension with \sys, and the remaining four were tied. This difference was significant among the eight non-tied participants (Wilcoxon signed-rank test, $p=.0078$). 


We also confirmed that the higher idea ratings could not be explained by differences in the number of ideas produced in each condition. Participants explored a similar number of ideas on the canvas in \sys ($M=33.8$, $SD=8.9$) and the baseline ($M=26.3$, $SD=7.8$), with no significant difference between conditions. They also submitted a similar number of ideas: an average of 4.83 ideas ($SD=1.27$) in \sys and 4.42 ideas ($SD=1.38$) in the baseline (Wilcoxon signed-rank test, $W=8.5$, $p=.347$).

\begin{table*}[t]
\caption{The examples of top and bottom scoring ideas from each condition in the user study.
``Cond.'' indicates the system condition; ``Prob.'' indicates the design task
(2: cooking support for older adults with cognitive decline; 
3: accessible communication of visual art style).
``Score'' represents the composite expert rating.
Idea texts are summarized for readability.}
\label{tab:high_low_idea_examples}
\centering
\renewcommand{\arraystretch}{1.1}
\begin{tabular}{p{0.08\textwidth} p{0.12\textwidth} p{0.07\textwidth} p{0.08\textwidth} p{0.59\textwidth}}
\toprule
\textbf{Rank} &
\textbf{Cond.} &
\textbf{Prob.} &
\textbf{Score} &
\textbf{Idea Example (Summarized)} \\
\midrule

Top &
\sys &
3 &
$>6$ &
A refreshable pin array normally presents a painting as a static tactile image. In an active mode, the same pins rise in the order the painting was created, letting visitors experience how the underpainting, major forms, and final details emerged over time from the artist's perspective.
\\
\addlinespace[0.5em]

Bottom &
\sys &
2 &
$<3$ &
Sensitive smoke detectors and heat sensors near the stove detect possible dangers and remind users.
\\

\midrule

Top &
Baseline &
2 &
$>6$ &
A divided ingredient tray lights the next compartment and shows one instruction at a time. Emptying a compartment marks the ingredient as added, starts the timer, and records the current step so the user can resume correctly after an interruption.
\\
\addlinespace[0.5em]

Bottom &
Baseline &
3 &
$<3$ &
A tablet lets visitors feel a tactile rendering of the artwork while listening to a paired song that conveys its visual style.
\\

\bottomrule
\end{tabular}
\end{table*}

\subsubsection{Comparing User generated ideas with ideas from Auto-Generation}

\paragraph{\textbf{\sys led to more highly rated ideas and greater variation in quality.}}
Ideas produced with \sys achieved overall higher expert ratings, comparable to those from the full auto-generation pipeline, while producing a larger proportion of highly rated ideas on several dimensions. 
On the same two tasks, ratings for the \sys condition and the full pipeline were closely matched on all four dimensions: Goal fulfillment (\sys: $M=5.16$,
$SD=0.99$; Auto: $M=5.28$, $SD=0.64$), New benefits (\sys:
$M=5.06$, $SD=0.97$; Auto: $M=5.18$, $SD=0.83$), Value vs. Cost
(\sys: $M=4.45$, $SD=1.22$; Auto: $M=4.98$, $SD=0.95$), and
Interestingness (\sys: $M=5.03$, $SD=1.17$; Auto: $M=5.05$,
$SD=0.71$), as well as on the composite score (\sys: $M=4.93$, $SD=0.76$;
Auto: $M=5.12$, $SD=0.54$). None of the differences was significant
(two-sided Mann--Whitney $U$ tests over the pooled ideas of both tasks; all
$p \geq .066$).

\begin{figure}[ht]
  \centering
 \includegraphics[width=0.8\linewidth]{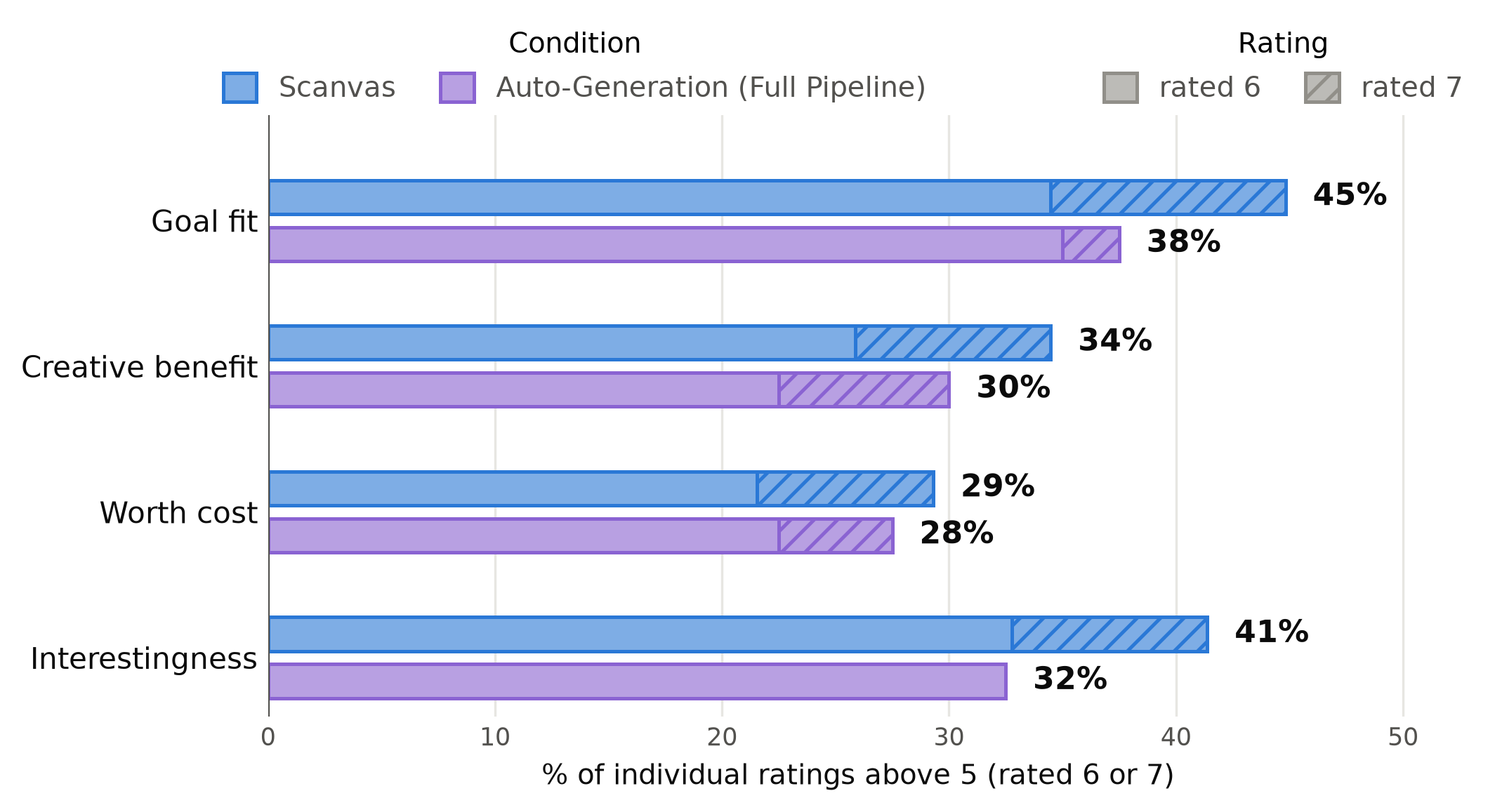}
  \caption{Percentage of high ratings (6 or 7 on a 7-point scale) per dimension for ideas from auto-generation and user sessions of \sys condition. Ideas from users working with \sys (n = 116 ratings) received a numerically higher share of top ratings than fully auto-generated ideas (n = 40 ratings) on all four dimensions.}
  \label{fig:high_end_ratings_diverging}
  \Description{Bar chart comparing the percentages of highly rated ideas from \sys auto-generation and \sys user sessions. In each of the four dimensions, a higher percentage of ideas from the user sessions received higher ratings. The exact percentages are described in the main text.}
\end{figure}

However, as shown in Figure~\ref{fig:high_end_ratings_diverging}, \sys had a larger proportion of high ratings (6--7) for Goal Fit (45\% vs.\ 38\%), Creative Benefit (34\% vs.\ 30\%), and Interestingness (41\% vs.\ 32\%), while Worth Cost was similar (29\% vs.\ 28\%). This pattern suggests that \sys supported participants in producing more ideas that reached the high end of the rating scale, despite similar overall ratings.
Overall, \sys produced a wider range of idea quality, including a small number of low-scoring ideas. Of the 58 ideas produced with \sys, two received ratings of 1--2 from both raters on at least one dimension, and another 10 received such a low rating from one rater. In contrast, all idea ratings for auto-generation full-pipeline were above 2. Participant interviews suggest that the limited study time sometimes constrained how fully participants could explore and describe their ideas, which may have contributed to these lower ratings. 

\paragraph{\textbf{\sys led to broader idea exploration than auto-generation.}}
As shown in Table~\ref{table:idea-semantic-spread}, ideas produced with \sys were significantly more semantically diverse than those from the auto-generation conditions. This pattern suggests that incorporating participants' own ideas and decisions broadened the explored idea space beyond what was achieved through fully automated generation. 
Importantly, \sys preserved this diversity advantage of human involvement, while also producing substantially higher-rated ideas than the user-study baseline. These results suggest that \sys maintained the breadth introduced by human exploration, and helped participants better leverage AI's generative capabilities to develop that broad exploration into higher-quality ideas.

\begin{table}[!ht]
\centering
\small
\renewcommand{\arraystretch}{1.15}
\setlength{\tabcolsep}{6pt}
\begin{tabular}{lccc}
\toprule
\textbf{Condition} & \textbf{$n$ ideas} & \textbf{Semantic spread} & \textbf{95\% CI} \\
\midrule

\multicolumn{4}{l}{\textit{User study}} \\
\sys       & 58 & 0.564             & [0.513, 0.595] \\
Baseline   & 53 & 0.534             & [0.477, 0.559] \\

\addlinespace[2pt]
\multicolumn{4}{l}{\textit{Automated generation}} \\
Full pipeline   & 20 & 0.488$^{*}$   & [0.453, 0.524] \\
w/o synergy     & 20 & 0.425$^{***}$ & [0.390, 0.461] \\
w/o expansion   & 20 & 0.388$^{***}$ & [0.351, 0.421] \\

\bottomrule
\end{tabular}
\caption{Average pairwise embedding distance among ideas generated under each condition on the two tasks shared by the user study and automated-generation evaluation. \sys has greater semantic diversity among ideas than the automated generation conditions. Brackets report 95\% bootstrap confidence intervals, resampling participants for the user-study conditions and ideas for the automated conditions. Stars indicate significant differences from \sys after Holm correction across the four comparisons ($^{*}p<.05$, $^{**}p<.01$, $^{***}p<.001$).}
\label{table:idea-semantic-spread}
\end{table}


\begin{figure}[ht]
  \centering
 \includegraphics[width=\linewidth]{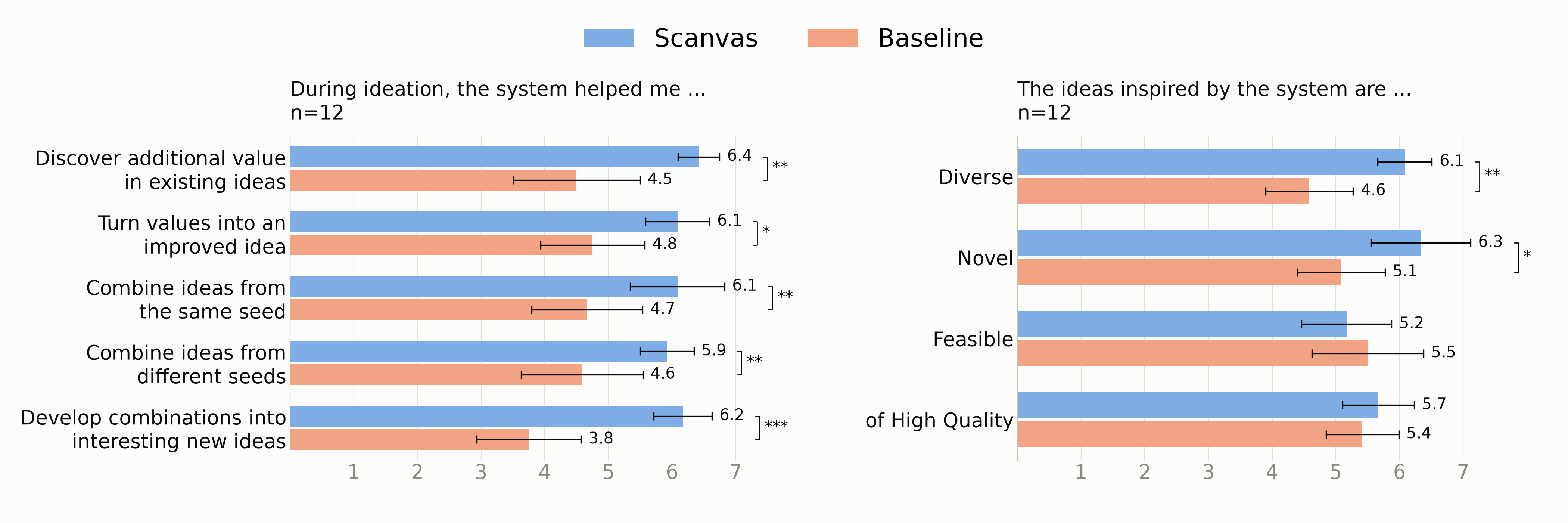}
  \caption{Participants' rating of their perceived experiences and the produced outcomes from the session. (Left) The graph shows that participants rated the their created ideas out of two conditions, the full system is significantly higher than the ablation baseline. (Right) The bar graph shows that ideas inspired by \sys were rated significantly more diverse and novel than those from the Baseline. Bars show means with 95\% CIs; two-sided paired-samples t-tests, *p<.05, **p<.01, ***p<.001.}
  \label{fig:user_rating_process}
  \Description{Bar charts comparing participants' subjective ratings of how much the system helped and how they would characterize the idea outputs in the two user study conditions. The key dimensions and results are described in the captions and main text.}
\end{figure}


\begin{figure}[ht]
  \centering
 \includegraphics[width=\linewidth]{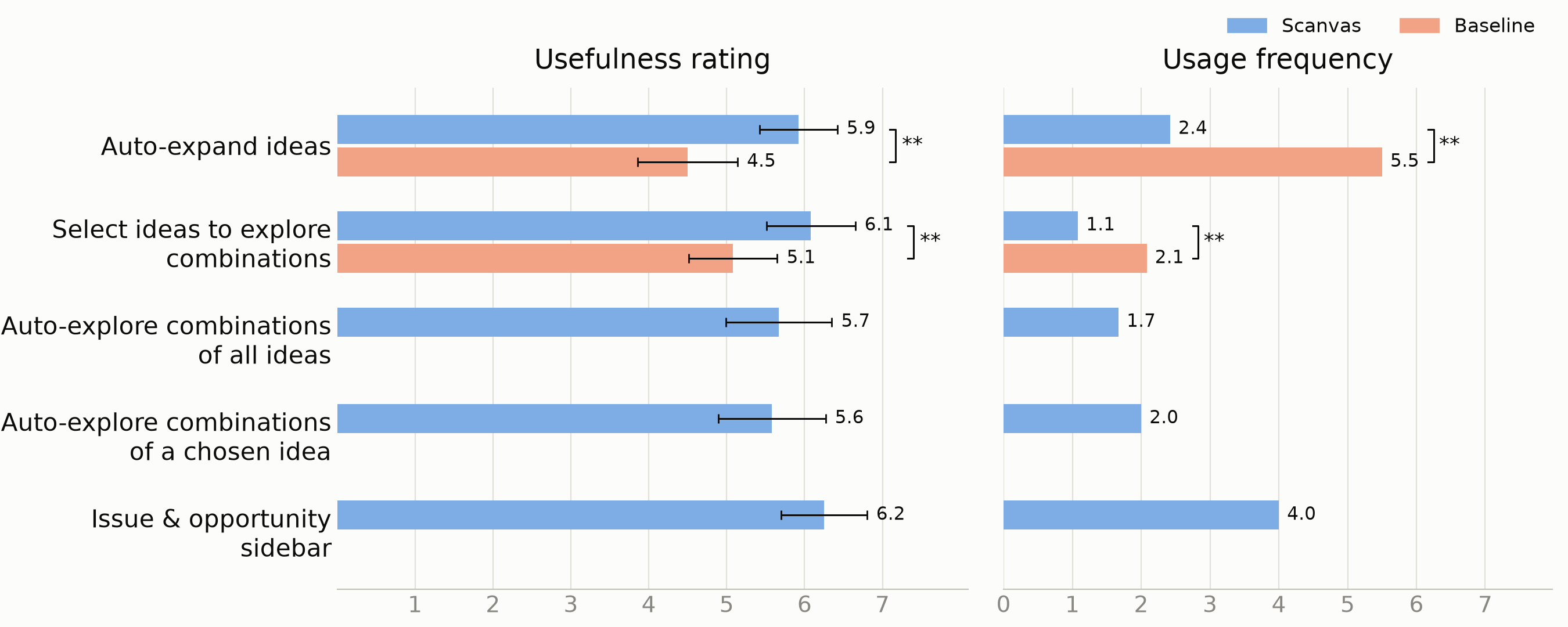}
\caption{Perceived usefulness and usage frequency of \sys features (n=12).
Left: post-task usefulness ratings on a 7-point Likert scale (higher = more useful);
bars show means, error bars 95\% confidence intervals. Right: the average count of each feature's usage frequency. Baseline bars are absent for features available only in
the full system. Brackets mark significant differences between conditions (paired
$t$-test; $^{*}p<.05$, $^{**}p<.01$, $^{***}p<.001$).}
  \label{fig:usefulness}
  \Description{Bar charts comparing participants' perceived usefulness and usage frequency of \sys features. The key dimensions and results are described in the captions and main text.}
\end{figure}

\subsubsection{User Creativity Measurement and Interaction Effects}
To test whether the results were largely influenced by participants' individual ability, we used the commonly adopted Alternate Uses Test (AUT)~\cite{gilhooly2007divergent} to assess participants' creativity, and examined the relationship between their AUT scores and the quality scores of the ideas they created (see Appendix~\ref{apd:pre-study} for details). We measured fluency as the total number of distinct alternative uses generated across the five prompts. We used Spearman rank correlations to examine the relationship between AUT fluency and each participant's mean idea-quality ratings (z-scored within task), as well as their within-participant \sys-minus-Baseline gains. The results  show that no correlation reached significance (quality: $|\rho| \leq .45$, all $p \geq .14$; gains: $|\rho| \leq .35$, all $p \geq .27$). These results suggest that \sys's benefits did not depend on participants' pre-existing divergent-thinking ability.

\subsection{User Feedback on How Enrichment Supported Synergy Discovery}
Participants perceived \sys as more helpful than the baseline for discovering additional value in existing ideas and developing that value into improved ideas. Comparing to baseline, they gave significantly higher ratings to these two aspects of the ideation process (Figure~\ref{fig:user_rating_process} left), which closely relate to the idea analysis and expansion features. These features are designed to enrich idea space to scaffold later stage synergy generation, and they received relatively high usefulness ratings (Figure~\ref{fig:usefulness} left). Participants' comments during the study session further illustrated how enrichment provided material for synergy discovery and supported further exploration after ideas had been combined.

The expansion and idea analysis features helped surface new properties and directions that created material for synergy discovery. Participants described the sidebar analysis as exposing opportunities and issues that they could use to guide further exploration. As P12 reflected, \pquote{(the side bar information) helped me identify what I like and what I don't like. I think that was a big pro...that helped me come up with better ideas because now I have a better understanding of what I want and don't want in ideas. } \sys's expansions complemented the sidebar analysis by offering diverse directions that remained connected to the original ideas. For example, in the task of designing kitchen support for older adults, P4 proposed a robot that would cook for the user. The expansion gave the robot an additional role: tracking missing ingredients and reminding the user what to buy. The sidebar highlighted that having the robot cook could reduce the user's enjoyment of cooking and opportunities to practice. Meanwhile, the analysis of a recipe-card idea indicated that it could require too much reading. The two ideas also represented opposite levels of assistance: the recipe card provided too little support, whereas the robot risked taking over the task and removing the user's enjoyment and opportunities to practice. These details helped P4 recognize a connection that was not clear from the original ideas: the robot's reminder capabilities could be combined with the portable card format to provide a more balanced level of assistance. P4 selected the ideas using \textit{+Mix}, producing an assistant that provided reminders and timing support while the user remained the cook and turned its ingredient knowledge into simple printed or digital shopping cards for use outside the home. P4 highlighted the value of the sidebar analysis: \pquote{I was most wowed by the (opportunity) options}. In this case, the expansion introduced a reusable capability, while the sidebar exposed issues that guided how this capability could be combined with another idea to support both independent cooking and shopping.

Enrichment could also continue after combination by surfacing further opportunities in the combined idea. For example, P11 explored a combination that added headphones to a tactile-card interaction. Analysis of the combined idea surfaced an opportunity to record visitors' finger movements on the tactile interface and use those data to improve the museum's audio descriptions. The movement patterns could indicate which parts of an artwork visitors found interesting or confusing. Here, the combined idea became material for another round of enrichment, which suggested a further connection between touch and audio. This illustrates an iterative workflow in which participants could revisit a combination's properties and continue developing it.

The comparison with the baseline suggests that enrichment supported a more deliberate process of interpreting and developing ideas by surfacing useful directions that participants could spend time examining, rather than simply encouraging them to generate more variants. In baseline condition, participants found the expansion feature less useful for further exploration. P5, P10, P11, and P12 described these expansions as elaborations of the original ideas or as ideas that remained very similar to them. P2 also described some expansions as \pquote{out of scope} or \pquote{not interesting}. These accounts help explain the lower perceived usefulness of baseline expansion (Figure~\ref{fig:usefulness}, left). More frequent use was therefore not necessarily the same as more useful support. As shown in Figure~\ref{fig:usefulness} (right), participants used the baseline expansion and idea-selection features more frequently despite rating them as less useful. Together with participants' comments, this pattern suggests that repeated baseline use may partly reflect trial-and-error, with participants repeatedly clicking to generate or combine ideas in search of a useful direction. In \sys, by contrast, enrichment surfaced potentially useful properties and directions for participants to examine, giving them more material to interpret and develop before deciding what to expand or combine next. Thus, participants could spend more of their interaction engaging with and developing promising ideas, rather than repeatedly invoking generation features to search for one.

\subsection{User Experience of Synergy Discovery}
\begin{figure}[ht]
  \centering
 \includegraphics[width=0.6\linewidth]{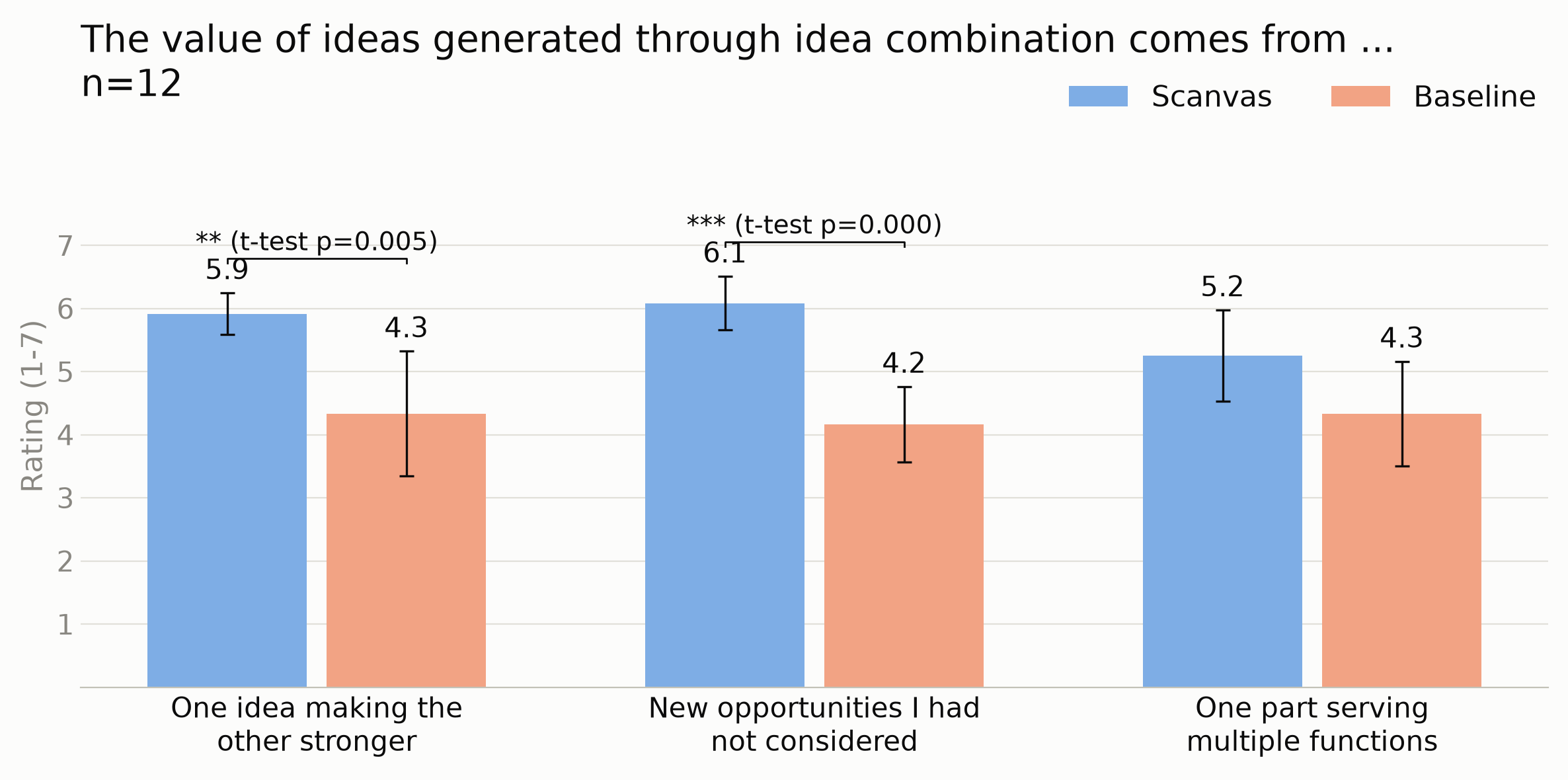}
  \caption{Participants' ratings of their perceived value of the ideas generated from the combine features in both conditions. Bars show means with 95\% CIs; n=12; *p<.05, **p<.01, ***p<.001.}
  \label{fig:user_rating_combine}
  \Description{Bar charts comparing participants' ratings of heir perceived value of the ideas generated from the combine features in the two user study conditions. \sys was rated to be superior in creating synergies in terms of strengthening ideas and revealing new opportunities.}
\end{figure}

Overall, participants perceived \sys as more helpful than the baseline for finding and developing useful combinations among ideas, as shown in Figure~\ref{fig:user_rating_process}(left), Figure~\ref{fig:usefulness}(left). 
Their experiences suggest that \sys supported synergy discovery in two connected steps: finding relationships that were hard to notice from individual ideas, and developing those relationships to create new value or achieve value with fewer resources or costs. The interviews further show how the three synergy strategies supported these steps.

\subsubsection{Helping users find promising idea combinations they might otherwise miss.}
Participants did not always recognize which existing idea pairs worth combining. The automatic combination features therefore helped participants explore pairings they might not have selected themselves.

For example, in the design task of support older adults in kitchen, P7 had separately explored a safety knob that could shut off the stove when it detected smoke and a screen display that guided an older adult through cooking steps. P7 saw them as separate tools and had not considered combining them. However, \sys's auto synergy discovery combined the two into a cooking practice mode: the burner stayed off while the knob still turned and clicked as in real cooking, and the display guided users through each step. This allowed older adults to rehearse the cooking process safely and build confidence before using the same knob and display during real cooking. P7 reflected, \pquote{I didn't realize if I could combine those two, that that could give me what I actually thought about.} They further explained, \pquote{when we combined safety knob and then this display I got this really nice practice mode that has both things, but I didn't think that you can combine them. I just thought they would be separate.} Here, \sys surfaced a promising relationship already present in the idea space but not apparent when the ideas were considered separately.

\subsubsection{Combining ideas to unlock new value that neither idea provided alone.}
Participants reported \sys helped them discover new values beyond what either parent idea provided, which most directly reflected the \emph{Combine to Unlock New Goals} strategy incorporated in \sys.

For example, P11 explored two ways to communicate a painting's visual style to visitors with visual impairments. One idea used a tactile card to communicate properties such as color and composition. \sys further developed the card by adding a compass so that it could identify which painting the visitor was facing. Another idea represented the painting through sound, which visitors could listen to through headphones. \sys combined the two ideas to create a new `shared experience': when visitors and their companions used the orientation-aware cards, the system could identify which part of the artwork each of them were attending to, and use a shared 3D sound space to transform the specific part of art into sound. P11 emphasized, \pquote{I really think that this is the biggest thing that task two had that was really nice... you're finding these new opportunities, especially ones that I didn't consider. I think I was close to considering [it], but I didn't fully consider that opportunity.} The combination did more than merge tactile and audio access methods: it unlocked a new goal of shared social experience that neither parent idea provided alone. In contrast, many combinations produced by the baseline system were described by participants, including P11, as reasonable elaborations but often remained close to the feature additions they had expected.

\subsubsection{Combining ideas to reduce the cost or resources required by the original ideas.}
Participants also valued combinations that achieved the original ideas' goals with fewer resources or addressed their limitations without introducing substantial additional resource demands. These cases reflected the \emph{Turn Weaknesses into Resources} and \emph{Share Components Across Functions} strategies incorporated in \sys.

For example, P2 explored one idea that required a dedicated tactile card for each artwork and another that communicated visual properties through vibration. P2 expected the combination to keep both, using tactile cards together with a vibration wristband. However, producing a custom card for every artwork was costly. Instead of adding vibration to the cards, \sys moved the interaction onto a phone the visitor already carries. Vibration patterns and guided finger gestures communicated each painting's visual style, allowing one device to support many artworks without separate tactile cards. P2 was happy to see this synergy: \pquote{you needed different kind of cards for every artwork. So surprisingly it also mitigated that when I never asked for it... there was no phone involved in [either parent].} Here, the weakness of the costly cards drove a combination that preserved the value while removing the resource-intensive component.

Participants also valued combinations in which one component could serve multiple functions to reduce the resources required. P10 had explored a spoon rest clipped to a pot that could chime or light up to prompt the next cooking step, as well as a separate hood-mounted sensor that monitored the stove and warned when a burner was left on. Rather than retaining separate devices for prompting and sensing, \sys used the spoon rest for both functions. Because it was already in contact with the pot, it could sense heat and vibration associated with cooking while also providing prompts, making the hood-mounted sensor only an optional backup. P12 described this pattern as \pquote{the same component but serving different [functions]... it helped combine ideas in a way that would increase the versatility of an idea.}

These cases show how \sys supported combinations that created additional value without proportionally adding components or resources. Rather than simply accumulating the features of the original ideas, the combinations produced by \sys addressed their limitations or reused existing components across functions.
\section{Discussion}
\subsection{From Generating Ideas to Developing Synergistic Combinations}

Our findings point to creating synergistic idea combinations as an important direction for expanding computational support for idea generation beyond searching for external inspiration. Much prior work in creativity support has focused on retrieving, filtering, and presenting external examples to help people reach new parts of a design space~\cite{luo2021guiding,wang2024scimon}, but the space available through retrieval is ultimately bounded by what can be found and recognized as relevant. Another source of new ideas lies within the existing idea space itself: existing ideas can be developed and recombined~\cite{gu2024llms,wang2023popblends} so that previously hidden relationships become useful new directions. This perspective is consistent with theories of recombinant innovation, which describe innovation as arising from new combinations of existing components and knowledge~\cite{fleming2001recombinant,weitzman1998recombinant}. \sys extends this view by treating combination not as simply placing two ideas together, but as a search for relationships that can create greater value from the same or fewer resources. Our results suggest that systematically surfacing idea properties and developing them before searching for these relationships can make the existing idea space itself a productive source of new ideas, complementing approaches that primarily search outward for additional inspiration.

Importantly, the value of searching through combinations may lie not in consistently producing better ideas, but in expanding the range of possible outcomes. Prior work on recombinant technological search argues that people's experimentation with new combinations can produce less successful outcomes while also increasing the variability needed for breakthrough inventions~\cite{fleming2001recombinant}. This suggests that variability can be a productive feature of creative search: exploring new relationships among existing ideas may open directions that incremental development would not. Creativity support systems might therefore support exploration of uncertain but potentially high-value combinations, rather than optimizing toward consistently high-quality outputs.

\subsection{Supporting Human--AI Collaboration Through Exploration and Pivoting}
Our results suggest that incorporating human judgment during idea-space exploration can create a productive interaction with AI search, helping promising directions emerge and develop into higher-end ideas. Compared with the end-to-end automated pipeline, the interactive condition produced a larger share of highly rated ideas across several evaluation dimensions. One possible explanation is that people can recognize latent potential as ideas develop, and pivot toward directions that automated search may not prioritize. These interventions then change what the AI expands and combines next. In contrast, an automated pipeline can systematically search a much larger combinatorial space and rapidly develop many possible relationships without requiring continuous user effort. This interaction extends our finding that human--AI ideation can combine human-driven exploration with AI-supported exploitation~\cite{li2026human}. People can introduce broader or more meaningful directions and judge which emerging ideas are worth pursuing, while LLMs can rapidly develop the local space around those pivots, and search for relationships with other ideas.

For synergy discovery, future systems could therefore combine the scale and coverage of automated combinatorial search with human intervention that continually reshapes the search as new potential becomes visible. Doing so will also require better ways to organize the growing idea space~\cite{sarica2024innovation} so people can better understand and explore where the search has already been done.

\subsection{Toward Value Functions for Evaluating Synergistic Ideas}
Our definition of synergy suggests a way to help people navigate increasingly large AI-generated idea spaces by identifying which possibilities may be worth further attention. Idea quality cannot be determined by novelty alone: creativity is commonly understood to also require usefulness, while design evaluation further considers whether ideas are feasible and workable~\cite{runco2012standard,dean2006identifying}. \sys frames synergy as creating greater value from the same or fewer resources, suggesting a value function that considers both the benefits an idea creates and the costs or resources required to realize them. Such estimates could help people filter and prioritize a large space of generated possibilities while retaining human judgment over what is worth pursuing. This could extend prior creativity-support approaches for organizing and filtering large idea spaces~\cite{siangliulue2016ideahound,lykourentzou2018crowds}. Future work could investigate how human judgment, domain knowledge, and computational models can jointly estimate idea value and cost computationally to help people navigate and focus their exploration.



\subsection{Limitations and Future Work}

Our study also points to several limitations and opportunities for future work. First, as participants generated and expanded more ideas, the canvas could become increasingly difficult to navigate. \sys currently provides limited support for organizing a growing idea space, which may make it harder for users to compare related ideas, revisit promising directions, or identify where further exploration is most useful. Future systems could investigate clustering, filtering, summarization, or adaptive views that help users browse and manage larger collections of ideas without losing the relationships among them.
Second, our evaluation captured only a short period of design activity under controlled task settings. Although the tasks were informed by professional design problems, participants worked within predefined scenarios and limited study time rather than pursuing ideas as part of an ongoing project. How designers use idea expansion and synergy discovery may change when they face real project constraints, accumulated domain knowledge, collaboration, and longer cycles of refinement. Longer-term deployments in active design projects, or broader public use of the system, could help reveal how these forms of exploration fit into everyday design practice and where additional support is needed.
Finally, our participant sample was relatively small and cannot capture the full range of design expertise, domains, and working styles found in professional practice. Different designers may vary in how they interpret surfaced opportunities, decide which ideas to pursue, and benefit from AI support. Future studies with a larger and more diverse set of practitioners could examine how these effects differ across levels of expertise, design domains, and individual approaches to ideation.

\subsection{Conclusion}
In this work, we present \sys, an AI-supported ideation approach for discovering and developing synergistic opportunities in design spaces. By analyzing and enriching existing ideas before intentionally searching for relationships among them, \sys helps users move beyond generating isolated variations or simply combining features toward creating additional value through how ideas interact. Across an end-to-end pipeline evaluation and a user study with design professionals, our findings show that this approach can improve idea quality and help surface promising relationships and design directions that might otherwise be overlooked.  We envision future creativity support systems that treat relationships among ideas as a central object of exploration, with AI broadening the space of possible connections while people guide which opportunities are worth developing.

\bibliographystyle{plainnat}
\bibliography{sample-base}

@article{suwa2000unexpected,
  author  = {Suwa, Masaki and Gero, John and Purcell, Terry},
  title   = {Unexpected Discoveries and S-Invention of Design Requirements: Important Vehicles for a Design Process},
  journal = {Design Studies},
  volume  = {21},
  number  = {6},
  pages   = {539--567},
  year    = {2000},
  doi     = {10.1016/S0142-694X(99)00034-4}
}

@article{weitzman1998recombinant,
  author  = {Weitzman, Martin L.},
  title   = {Recombinant Growth},
  journal = {The Quarterly Journal of Economics},
  volume  = {113},
  number  = {2},
  pages   = {331--360},
  year    = {1998},
  doi     = {10.1162/003355398555595}
}

@inproceedings{difede2022ideamachine,
  author    = {Di Fede, Giulia and Rocchesso, Davide and Dow, Steven P. and Andolina, Salvatore},
  title     = {The Idea Machine: LLM-based Expansion, Rewriting, Combination, and Suggestion of Ideas},
  booktitle = {Proceedings of the 14th Conference on Creativity and Cognition},
  pages     = {623--627},
  year      = {2022},
  publisher = {Association for Computing Machinery},
  doi       = {10.1145/3527927.3535197}
}

@inproceedings{suh2024luminate,
  author    = {Suh, Sangho and Chen, Meng and Min, Bryan and Li, Toby Jia-Jun and Xia, Haijun},
  title     = {Luminate: Structured Generation and Exploration of Design Space with Large Language Models for Human-AI Co-Creation},
  booktitle = {Proceedings of the 2024 CHI Conference on Human Factors in Computing Systems},
  year      = {2024},
  publisher = {Association for Computing Machinery},
  pages     = {1--26},
  doi       = {10.1145/3613904.3642400}
}

@inproceedings{wang2023popblends,
  author    = {Wang, Sitong and Petridis, Savvas and Kwon, Taeahn and Ma, Xiaojuan and Chilton, Lydia B.},
  title     = {PopBlends: Strategies for Conceptual Blending with Large Language Models},
  booktitle = {Proceedings of the 2023 CHI Conference on Human Factors in Computing Systems},
  articleno = {435},
  numpages  = {19},
  year      = {2023},
  publisher = {Association for Computing Machinery},
  doi       = {10.1145/3544548.3580948}
}

@article{chan2015importance,
  title={The importance of iteration in creative conceptual combination},
  author={Chan, Joel and Schunn, Christian D},
  journal={Cognition},
  volume={145},
  pages={104--115},
  year={2015},
  publisher={Elsevier},
  doi={10.1016/j.cognition.2015.08.008}
}

@article{wang2026nexusai,
  title={NexusAI: Enabling Design Space Exploration of Ideas through Cognitive Abstraction and Functional Decomposition},
  author={Wang, Anqi and Wang, Bingqian and Chen, Huiyang and Jiao, Keqing and Han, Lei and Tong, Xin and Hui, Pan},
  journal={arXiv preprint arXiv:2604.10575},
  year={2026}
}

@article{yang2025flexmind,
  title={FlexMind: Supporting Deeper Creative Thinking with LLMs},
  author={Yang, Yaqing and Mohanty, Vikram and Chen, Yan-Ying and Hong, Matthew K and Martelaro, Nikolas and Kittur, Aniket},
  journal={arXiv preprint arXiv:2509.21685},
  year={2025}
}

@inproceedings{wang2024scimon,
  title={Scimon: Scientific inspiration machines optimized for novelty},
  author={Wang, Qingyun and Downey, Doug and Ji, Heng and Hope, Tom},
  booktitle={Proceedings of the 62nd annual meeting of the association for computational linguistics (volume 1: long papers)},
  pages={279--299},
  year={2024}
}

@article{gu2024llms,
  title={LLMs can realize combinatorial creativity: generating creative ideas via LLMs for scientific research},
  author={Gu, Tianyang and Wang, Jingjin and Zhang, Zhihao and Li, HaoHong},
  journal={arXiv preprint arXiv:2412.14141},
  year={2024}
}

@article{chen2025structuring,
  title={Structuring scientific innovation: A framework for modeling and discovering impactful knowledge combinations},
  author={Chen, Junlan and Zhang, Kexin and Li, Daifeng and Feng, Yangyang and Zhang, Yuxuan and Deng, Bowen},
  journal={arXiv preprint arXiv:2503.18865},
  year={2025}
}

@article{koo2016guideline,
  title={A guideline of selecting and reporting intraclass correlation coefficients for reliability research},
  author={Koo, Terry K and Li, Mae Y},
  journal={Journal of chiropractic medicine},
  volume={15},
  number={2},
  pages={155--163},
  year={2016},
  publisher={National University of Health Sciences}
}

@article{boden1998creativity,
  title={Creativity and artificial intelligence},
  author={Boden, Margaret A},
  journal={Artificial intelligence},
  volume={103},
  number={1-2},
  pages={347--356},
  year={1998},
  publisher={Elsevier}
}

@inproceedings{pu2025ideasynth,
  title={Ideasynth: Iterative research idea development through evolving and composing idea facets with literature-grounded feedback},
  author={Pu, Kevin and Feng, KJ Kevin and Grossman, Tovi and Hope, Tom and Dalvi Mishra, Bhavana and Latzke, Matt and Bragg, Jonathan and Chang, Joseph Chee and Siangliulue, Pao},
  booktitle={Proceedings of the 2025 CHI Conference on Human Factors in Computing Systems},
  pages={1--31},
  year={2025}
}

@inproceedings{chen2024triz,
  title={TRIZ-GPT: An LLM-augmented method for problem-solving},
  author={Chen, Liuqing and Song, Yaxuan and Ding, Shixian and Sun, Lingyun and Childs, Peter and Zuo, Haoyu},
  booktitle={International Design Engineering Technical Conferences and Computers and Information in Engineering Conference},
  volume={88407},
  pages={V006T06A010},
  year={2024},
  organization={American Society of Mechanical Engineers}
}

@inproceedings{jiang2024autotriz,
  title={Autotriz: Artificial ideation with triz and large language models},
  author={Jiang, Shuo and Luo, Jianxi},
  booktitle={International Design Engineering Technical Conferences and Computers and Information in Engineering Conference},
  volume={88377},
  pages={V03BT03A055},
  year={2024},
  organization={American Society of Mechanical Engineers}
}

@inproceedings{sternlicht2026chimera,
  title={CHIMERA: a knowledge base of scientific idea recombinations for research analysis and ideation},
  author={Sternlicht, Noy and Hope, Tom},
  booktitle={Proceedings of the 64th Annual Meeting of the Association for Computational Linguistics (Volume 1: Long Papers)},
  pages={1871--1905},
  year={2026}
}

@article{gilhooly2007divergent,
  title={Divergent thinking: Strategies and executive involvement in generating novel uses for familiar objects},
  author={Gilhooly, Kenneth J and Fioratou, Evridiki and Anthony, Susan H and Wynn, Victor},
  journal={British Journal of Psychology},
  volume={98},
  number={4},
  pages={611--625},
  year={2007},
  publisher={Wiley Online Library}
}

@inproceedings{li2026human,
  title={Human-AI Synergy Supports Collective Creative Search},
  author={Li, Chenyi and Marjieh, Raja and Hu, Haoyu and Steyvers, Mark and Collins, Katherine M and Sucholutsky, Ilia and Jacoby, Nori},
  booktitle={Proceedings of the Annual Meeting of the Cognitive Science Society},
  volume={48},
  year={2026}
}

@article{wilkenfeld2001similarity,
  title={Similarity and emergence in conceptual combination},
  author={Wilkenfeld, Merryl J and Ward, Thomas B},
  journal={Journal of Memory and Language},
  volume={45},
  number={1},
  pages={21--38},
  year={2001},
  publisher={Elsevier}
}

@article{schorlemmer2014coinvent,
  title={Coinvent: Towards a computational concept invention theory},
  author={Schorlemmer, Marco and Smaill, Alan and K{\"u}hnberger, Kai-Uwe and Kutz, Oliver and Colton, Simon and Cambouropoulos, Emilios and Pease, Alison},
  year={2014}
}

@article{han2018combinator,
  title={The Combinator--a computer-based tool for creative idea generation based on a simulation approach},
  author={Han, Ji and Shi, Feng and Chen, Liuqing and Childs, Peter RN},
  journal={Design Science},
  volume={4},
  pages={e11},
  year={2018},
  publisher={Cambridge University Press}
}

@article{chen2019artificial,
  title={An artificial intelligence based data-driven approach for design ideation},
  author={Chen, Liuqing and Wang, Pan and Dong, Hao and Shi, Feng and Han, Ji and Guo, Yike and Childs, Peter RN and Xiao, Jun and Wu, Chao},
  journal={Journal of Visual Communication and Image Representation},
  volume={61},
  pages={10--22},
  year={2019},
  publisher={Elsevier}
}

@article{eppe2018computational,
  title={A computational framework for conceptual blending},
  author={Eppe, Manfred and Maclean, Ewen and Confalonieri, Roberto and Kutz, Oliver and Schorlemmer, Marco and Plaza, Enric and K{\"u}hnberger, Kai-Uwe},
  journal={Artificial Intelligence},
  volume={256},
  pages={105--129},
  year={2018},
  publisher={Elsevier}
}

@article{fleming2001recombinant,
  title={Recombinant uncertainty in technological search},
  author={Fleming, Lee},
  journal={Management science},
  volume={47},
  number={1},
  pages={117--132},
  year={2001},
  publisher={INFORMS}
}

@article{hatchuel2009ck,
  title={CK design theory: an advanced formulation},
  author={Hatchuel, Armand and Weil, Benoit},
  journal={Research in engineering design},
  volume={19},
  number={4},
  pages={181--192},
  year={2009},
  publisher={Springer}
}

@article{sarasvathy2001causation,
  title={Causation and effectuation: Toward a theoretical shift from economic inevitability to entrepreneurial contingency},
  author={Sarasvathy, Saras D},
  journal={Academy of management Review},
  volume={26},
  number={2},
  pages={243--263},
  year={2001},
  publisher={Academy of Management Briarcliff Manor, NY 10510}
}

@article{ilevbare2013review,
  title={A review of TRIZ, and its benefits and challenges in practice},
  author={Ilevbare, Imoh M and Probert, David and Phaal, Robert},
  journal={Technovation},
  volume={33},
  number={2-3},
  pages={30--37},
  year={2013},
  publisher={Elsevier}
}

@article{miron2011paradoxical,
  title={Paradoxical frames and creative sparks: Enhancing individual creativity through conflict and integration},
  author={Miron-Spektor, Ella and Gino, Francesca and Argote, Linda},
  journal={Organizational Behavior and Human Decision Processes},
  volume={116},
  number={2},
  pages={229--240},
  year={2011},
  publisher={Elsevier}
}

@article{ulrich1990function,
  title={Function sharing in mechanical design},
  author={Ulrich, Karl T and Seering, Warren P},
  journal={Design Studies},
  volume={11},
  number={4},
  pages={223--234},
  year={1990},
  publisher={Elsevier}
}

@book{fuller1982synergetics,
  title={Synergetics: explorations in the geometry of thinking},
  author={Fuller, R Buckminster},
  year={1982},
  publisher={Estate of R. Buckminster Fuller}
}

@article{braungart2007cradle,
  title={Cradle-to-cradle design: creating healthy emissions--a strategy for eco-effective product and system design},
  author={Braungart, Michael and McDonough, William and Bollinger, Andrew},
  journal={Journal of cleaner production},
  volume={15},
  number={13-14},
  pages={1337--1348},
  year={2007},
  publisher={Elsevier}
}

@article{chertow2000industrial,
  title={Industrial symbiosis: literature and taxonomy},
  author={Chertow, Marian R},
  journal={Annual review of energy and the environment},
  volume={25},
  number={1},
  pages={313--337},
  year={2000},
  publisher={Annual Reviews 4139 El Camino Way, PO Box 10139, Palo Alto, CA 94303-0139, USA}
}

@article{ulrich1995role,
  title={The role of product architecture in the manufacturing firm},
  author={Ulrich, Karl},
  journal={Research policy},
  volume={24},
  number={3},
  pages={419--440},
  year={1995},
  publisher={Elsevier}
}

@article{chakrabarti2004new,
  title={A new approach to structure sharing},
  author={Chakrabarti, Amaresh},
  journal={J. Comput. Inf. Sci. Eng.},
  volume={4},
  number={1},
  pages={11--19},
  year={2004}
}

@article{gero2004situated,
  title={The situated function--behaviour--structure framework},
  author={Gero, John S and Kannengiesser, Udo},
  journal={Design studies},
  volume={25},
  number={4},
  pages={373--391},
  year={2004},
  publisher={Elsevier}
}

@article{bikakis2021repurposing,
  title={Repurposing of resources: from everyday problem solving through to crisis management},
  author={Bikakis, Antonis and Dickens, Luke and Hunter, Anthony and Miller, Rob},
  journal={arXiv preprint arXiv:2109.08425},
  year={2021}
}

@article{chertow2007uncovering,
  title={“Uncovering” industrial symbiosis},
  author={Chertow, Marian R},
  journal={Journal of industrial Ecology},
  volume={11},
  number={1},
  pages={11--30},
  year={2007},
  publisher={Wiley Online Library}
}

@article{dorst2001creativity,
  title={Creativity in the design process: co-evolution of problem--solution},
  author={Dorst, Kees and Cross, Nigel},
  journal={Design studies},
  volume={22},
  number={5},
  pages={425--437},
  year={2001},
  publisher={Elsevier}
}

@article{cherry2014quantifying,
  title={Quantifying the creativity support of digital tools through the creativity support index},
  author={Cherry, Erin and Latulipe, Celine},
  journal={ACM Transactions on Computer-Human Interaction (TOCHI)},
  volume={21},
  number={4},
  pages={1--25},
  year={2014},
  publisher={ACM New York, NY, USA}
}

@misc{openai_text_embedding_3_small,
  author       = {{OpenAI}},
  title        = {text-embedding-3-small Model},
  howpublished = {\url{https://developers.openai.com/api/docs/models/text-embedding-3-small}},
  year         = {2026},
  note         = {Accessed: 2026-09-08}
}

@inproceedings{cox2021directed,
  title={Directed diversity: Leveraging language embedding distances for collective creativity in crowd ideation},
  author={Cox, Samuel Rhys and Wang, Yunlong and Abdul, Ashraf and Von Der Weth, Christian and Y. Lim, Brian},
  booktitle={Proceedings of the 2021 CHI Conference on Human Factors in Computing Systems},
  pages={1--35},
  year={2021}
}

@book{altshuller1984creativity,
  title={Creativity as an exact science},
  author={Altshuller, Genrikh Saulovich},
  year={1984},
  publisher={crc Press}
}

@article{luo2021guiding,
  title={Guiding data-driven design ideation by knowledge distance},
  author={Luo, Jianxi and Sarica, Serhad and Wood, Kristin L},
  journal={Knowledge-Based Systems},
  volume={218},
  pages={106873},
  year={2021},
  publisher={Elsevier}
}

@article{sarica2024innovation,
  title={The innovation paradox: concept space expansion with diminishing originality and the promise of creative artificial intelligence},
  author={Sarica, Serhad and Luo, Jianxi},
  journal={Design Science},
  volume={10},
  pages={e11},
  year={2024},
  publisher={Cambridge University Press}
}

@article{runco2012standard,
  title={The standard definition of creativity},
  author={Runco, Mark A and Jaeger, Garrett J},
  journal={Creativity research journal},
  volume={24},
  number={1},
  pages={92--96},
  year={2012},
  publisher={Taylor \& Francis}
}

@article{dean2006identifying,
  title={Identifying good ideas: constructs and scales for idea evaluation},
  author={Dean, Douglas L and Hender, Jill and Rodgers, Tom and Santanen, Eric},
  journal={Journal of Association for Information Systems},
  volume={7},
  number={10},
  pages={646--699},
  year={2006}
}

@inproceedings{siangliulue2016ideahound,
  title={IdeaHound: improving large-scale collaborative ideation with crowd-powered real-time semantic modeling},
  author={Siangliulue, Pao and Chan, Joel and Dow, Steven P and Gajos, Krzysztof Z},
  booktitle={Proceedings of the 29th Annual Symposium on User Interface Software and Technology},
  pages={609--624},
  year={2016}
}

@article{lykourentzou2018crowds,
  title={When crowds give you lemons: Filtering innovative ideas using a diverse-bag-of-lemons strategy},
  author={Lykourentzou, Ioanna and Ahmed, Faez and Papastathis, Costas and Sadien, Irwyn and Papangelis, Konstantinos},
  journal={Proceedings of the ACM on Human-Computer Interaction},
  volume={2},
  number={CSCW},
  pages={1--23},
  year={2018},
  publisher={ACM New York, NY, USA}
}

@article{duncker1945problem,
  title={On problem-solving.},
  author={Duncker, Karl and Lees, Lynne S},
  journal={Psychological monographs},
  volume={58},
  number={5},
  pages={i},
  year={1945},
  publisher={American Psychological Association}
}

@article{mccaffrey2012innovation,
  title={Innovation relies on the obscure: A key to overcoming the classic problem of functional fixedness},
  author={McCaffrey, Tony},
  journal={Psychological science},
  volume={23},
  number={3},
  pages={215--218},
  year={2012},
  publisher={Sage Publications Sage CA: Los Angeles, CA}
}

@article{jansson1991design,
  title={Design fixation},
  author={Jansson, David G and Smith, Steven M},
  journal={Design studies},
  volume={12},
  number={1},
  pages={3--11},
  year={1991},
  publisher={Elsevier}
}

@article{crilly2015fixation,
  title={Fixation and creativity in concept development: The attitudes and practices of expert designers},
  author={Crilly, Nathan},
  journal={Design studies},
  volume={38},
  pages={54--91},
  year={2015},
  publisher={Elsevier}
}

\section{Appendix}
\subsection{User Study Additional Details}

\subsubsection{Study System Tutorials}
\label{apd:tutorial}
Before starting each of the two main tasks, participants received a tutorial on the assigned system. The interviewer demonstrated the system step by step, introduced its main features, and asked participants to practice using it on a demo task: “Clean laundry with less water.” The tutorial for each condition took approximately 5 minutes.

\subsubsection{Post-study survey}
\label{apd:post-study}
After completing both system conditions, participants completed a post-study survey evaluating their experience with the two systems. The survey assessed participants' perceptions of the ideation process, the usefulness of individual system features, and the quality of the resulting ideas. We also administered the Creativity Support Index (CSI) to assess the systems' overall support for creative work. The corresponding ratings are reported in Table~\ref{table:csi}.

\begin{table}[!ht]
\centering
\small
\renewcommand{\arraystretch}{1.3}
\setlength{\tabcolsep}{5pt}
\begin{tabular}{p{0.62\linewidth} c c}
\hline
\textbf{Item} & \textbf{M} & \textbf{SD} \\
\hline
Results Worth Effort: I was satisfied with what I got out of the system or tool. & 6.08 & 0.90 \\
Expressiveness: I was able to be very creative while doing the activity inside this system. & 5.83 & 0.94 \\
Exploration: It was easy for me to explore many different ideas, options, designs,
or outcomes, using this system. & 5.75 & 1.14 \\
Immersion: My attention was fully tuned to the activity, and I forgot about the system  I was using. & 6.08 & 0.79 \\
Enjoyment: I would be happy to use this system or tool on a regular basis. & 6.42 & 0.79 \\
\hline
\end{tabular}
\caption{\textbf{Creativity Support Index items for the \sys condition} ($n=12$). Each item was rated post-task on a 7-point Likert scale (higher = better).}
\label{table:csi}
\end{table}

\subsubsection{Pre-study survey}
\label{apd:pre-study}
Before the study session, participants completed a pre-study survey covering demographic information, including age, sex, occupation, years of engineering experience, and domain of expertise. The survey also included an Alternative Uses Test (AUT) to characterize participants' baseline creative ability.  Each participant was presented with five objects in order: a coffee mug, tire, pair of pants, table, and bottle. They were asked to come up with alternative uses for each object within one minute.
Table~\ref{table:aut-fluency} reports the Spearman correlations between AUT fluency and each of the four idea-quality dimensions, both for pooled idea quality and for the within-participant \sys-minus-Baseline gain. It shows participants' pre-existing divergent-thinking ability (AUT fluency) did not predict how well they did in the study, nor how much \sys helped them.
We also collected informed consent from all participants before they began the study session.

\begin{table}[!ht]
\centering
\small
\renewcommand{\arraystretch}{1.3}
\setlength{\tabcolsep}{5pt}
\begin{tabular}{p{0.30\linewidth} r r r r}
\hline
 & \multicolumn{2}{c}{\textbf{Idea quality (pooled)}} & \multicolumn{2}{c}{\textbf{\sys$-$Baseline gain}} \\
\textbf{Rating dimension} & \multicolumn{1}{c}{$\rho$} & \multicolumn{1}{c}{$p$} & \multicolumn{1}{c}{$\rho$} & \multicolumn{1}{c}{$p$} \\
\hline
Goal fit          & $-.09$ & .79 & $.30$ & .34 \\
Creative benefit  & $-.05$ & .87 & $.13$ & .68 \\
Worth the cost    & $-.14$ & .67 & $.17$ & .59 \\
Interestingness   & $-.42$ & .18 & $.33$ & .30 \\
\hline
\end{tabular}
\caption{\textbf{Spearman correlations between AUT fluency and in-study idea quality} ($n=12$). AUT fluency is the total number of distinct alternative uses across the five prompts ($M=18.8$, $SD=5.5$). Idea quality (pooled) is each participant's mean idea rating (z-scored within task, pooled across conditions); gain is the within-participant \sys-minus-Baseline difference in those means. No correlation reached significance, and none survived Benjamini--Hochberg FDR correction within its test family (all $q \geq .65$). AUT fluency also did not predict the number of ideas produced in-study ($\rho=.27$, $p=.40$ for total ideas).}
\label{table:aut-fluency}
\end{table}
\end{document}